\documentclass[aps,prb,twocolumn,showpacs,superscriptaddress]{revtex4}

\usepackage{graphicx}
\usepackage{amsmath}

\newcommand{\bvec}[1]{{\bf #1}}

\newcommand{\bra}[1]{\left<#1 \right|} \newcommand{\ket}[1]{\left| #1
  \right>}

\newcommand{\Eref}[1]{Eq.~(\ref{#1})}

\begin{document}
 
\title{Correlated mesoscopic fluctuations in integer quantum Hall
transitions}

\author{Chenggang Zhou}

\affiliation{Department of Electrical Engineering, Princeton
  University, Princeton, New Jersey 08544, USA}

\altaffiliation{{\em Current address:} Computer Science and
Mathematics Devision, Oak Ridge National Laboratory, P.O. Box 2008, MS
6164, Oak Ridge Tennessee, 37831, USA; Center for Simulational
Physics, University of Georgia, Athens Georgia, 30602 USA.}

\author{Mona Berciu}

\affiliation{Department of Physics and Astronomy, University of
  British Columbia, Vancouver, BC V6T 1Z1, Canada } \date{\today}
 
\begin{abstract}

We investigate the origin of the resistance fluctuations of mesoscopic
samples, near transitions between Quantum Hall plateaus. These
fluctuations have been recently observed experimentally by E. Peled
{\em et al}. [Phys. Rev. Lett. {\bf 90}, 246802 (2003); {\em ibid}
{\bf 90}, 236802 (2003); Phys. Rev. B {\bf 69}, 241305(R) (2004)].  We
perform realistic first-principles simulations using a six-terminal
geometry and sample sizes similar to those of real devices, to model
the actual experiment.  We present the theory and implementation of
these simulations, which are based on the linear response theory for
non-interacting electrons.  The Hall and longitudinal resistances
extracted from the Landauer formula exhibit all the observed
experimental features.  We give a unified explanation for the three
regimes with distinct types of fluctuations observed experimentally,
based on a simple generalization of the Landauer-B\"uttiker model. The
transport is shown to be determined by the interplay between tunneling
and chiral currents. We identify the central part of
the transition, at intermediate filling factors, as the critical
region where the localization length is larger than the sample size.

\end{abstract}


\maketitle

\section{Introduction}
\label{sintro}

The integer quantum Hall effect\citep{Klitzing80} (IQHE) is one of
very few instances where quantum effects are dramatically manifested
in the {\em macroscopic} world: if a two-dimensional electron system
(2DES) is placed in a large perpendicular magnetic field $B$, its Hall
resistance is quantized with high accuracy to $h/(ne^2)$, where $n>0$
is an integer. This quantization has been explained, within the
non-interacting electron approximation, as a bulk effect,
\citep{Streda82a} as a topological invariant,\citep{TKNM82} and as an
edge effect. \citep{Rammal83,Buttiker88} The plateau-to-plateau
transitions of the Hall resistance, which are accompanied by peaks in
the longitudinal resistance, are understood as
localization-delocalization transitions with a universal scaling
relation.\citep{Pruisken87,Tsui88,Huckestein90,Bhatt92,Huckestein95}

A different route to observing quantum effects in condensed matter
physics, is to reduce the size of the device to the so-called
{\em mesoscopic} regime.  If the  size of the device is
smaller than the  phase coherence length, quantum interference
between different paths leads to effects such as random universal
conductance fluctuations.\cite{Beenakker97,Altshuler91,Lee87}

What happens if the two are combined?  The answer was provided by
recent IQHE experiments performed on a mesoscopic
sample.\citep{Peled03a,Peled03b,Peled04} The expected quantum
mechanical interference of electronic wave-functions indeed causes
reproducible patterns of fluctuations in the resistances near
plateau-to-plateau transitions, as the magnetic field is
varied  (in macroscopic samples all resistances vary
smoothly with $B$). Interestingly, the fluctuations of resistances
measured with 
different combinations of electrical contacts are not independent;
instead, various non-trivial correlations were observed.

In this paper, we investigate the mesoscopic IQHE within the
 non-interacting electron approximation. Our numerical simulations
 provide the basis for a unified explanation of these novel
 experimental observations, based on existing theories of the IQHE and
 mesoscopic transport (some of these results have been summarized in
 Ref.~\onlinecite{Zhou04c}). We begin the paper with a 
 brief review of the experimental observations, in Sec.~\ref{sec2}.
 In Sec.~\ref{numsim} we describe in detail the model we use, and our
 approach to solve the scattering problem that allows us to compute
 the conductance matrix from the multi-terminal Landauer formula.  Our
 numerical results, which exhibit all the experimentally observed
 symmetries and correlations, are presented in Sec.~\ref{numres}.  In
 Sec.~\ref{sglparam} we explain how these various correlations and
 symmetries arise. We analyze the allowed structure of the conductance
 matrix, and show how it directly determines the possible
 correlations between fluctuations of various resistances. Finally,
 Sec.~\ref{ssum} contains the summary and a discussion of the relevance
 of these results.

\section{Summary of Experimental results}
\label{sec2}

\begin{figure}[t]
\centering \includegraphics[width=0.95\columnwidth]{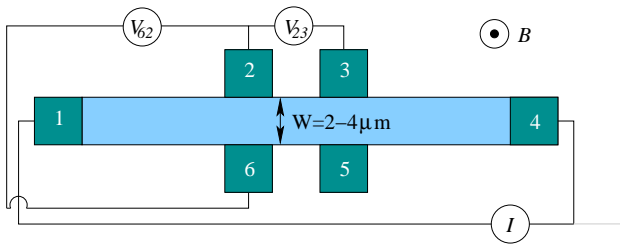}
\caption{(color online) Typical six-terminal Hall bar used in experiments. In this
configuration, the Hall resistance $R^H_{14,62}$ and the longitudinal
resistance $R^L_{14,23}$ are measured.}
\label{fig1}
\end{figure}

The experimental sample is an InGaAs/InAlAs heterostructure with
impurities (Indium atoms) distributed throughout.  As a result, the
mobility is rather low (fractional quantum Hall effect is not
observed). Even for the IQHE, only the first few plateau-to-plateau
transitions can be identified.  The experimental setup is sketched in
Fig.~\ref{fig1}. The Hall bar is connected to six terminals, indexed 1
to 6. The sample is considered mesoscopic because the characteristic
size of its central region between the 4 central terminals, of about
$2\mu$m, is comparable to the phase coherent length
$L_\phi\approx2-3\mu$m.\citep{Peled03a,Peled03b, Peled04}

As customary, we denote the various resistances as $R_{ij,mn} =
V_{mn}/ I_{ij}$, where $V_{mn}=V_n-V_m$ is the voltage difference
measured between terminals $n$ and $m$, when the current is injected in
the sample through terminal $i$ and extracted through terminal
$j$. For example, to measure the Hall and longitudinal resistances, a
small current $I$ is fed into terminal 1 and collected from terminal
4; no net currents are flowing in the other 4 terminals. One can then
measure two longitudinal resistances $R^L_{14,23}$, $R^L_{14,65}$, and
two Hall resistances $R^H_{14,62}$, $R^H_{14,53}$.  For any
macroscopic sample, the two resistances of each pair are equal. For a
mesoscopic sample, however, each resistance exhibits a different
fluctuation pattern.\citep{Peled03a,Peled03b, Peled04}

Another quantity of interest is the ``two-terminal resistance''
$R_{2t} = R_{63,63}$. Experimentally, it is measured by sending the
current from terminal 6 to terminal 3, and measuring the voltage
difference between terminal 6 and terminal 3. Because of the two-point
measurement, a contact resistance is subtracted from the raw
data.\cite{Peled03b}

Summary of experimental findings:

A: the IQHE transition from the $n^{th}$ to the $n+1^{st}$ plateau,
$\forall n>0$, has three distinct regimes:\citep{Peled03b,Peled04}

A1: on the high-$B$ (low filling factor $ \nu$) side of the
transition, the Hall and the longitudinal resistances have distinct
fluctuation patterns, which however are correlated such that:
\begin{equation}
\label{e1.1}
R^H_{14,62}+R^L_{14,23}=R^H_{14,53}+R^L_{14,65}= {h \over ne^2}
\end{equation}

A2: in the central part of the transition, the Hall and the
longitudinal resistances still exhibit distinct fluctuation patterns,
but Eq.~(\ref{e1.1}) is no longer satisfied.

A3: on the low-$B$ (high filling factor $ \nu$) side of the
transition, the Hall resistances become quantized to the expected
plateau value $R^H_{14,62}=R^H_{14,53}=h/[(n+1)e^2]$. The two
longitudinal resistances exhibit significant, but this time {\em
identical} fluctuation patterns, with a maximum amplitude
$h/[n(n+1)e^2]$, i.e.  the difference between quantized Hall
resistances of the two neighboring plateaus.

B: the transition inside the lowest Landau level $n=0$ shows the
equivalent of A3: at high-$\nu$ one finds the regime where
$R^H_{14,62}=R^H_{14,53}=h/e^2$, 
while the two longitudinal resistances have identical fluctuation
patterns with extremely large amplitudes.\citep{Peled03a} 
Regimes A1 and A2 are replaced here by the transition to the insulating
phase, where all four resistances increase rapidly with decreasing
filling factor.

C: throughout each IQHE transition, the following identity is found to
hold to high accuracy:\citep{Peled03b}
\begin{equation}
\label{e1.2} 
R^H_{14,62}+R^L_{14,23}=R^H_{14,53}+R^L_{14,65}=R_{63,63}.
\end{equation}

D: Under the reversal of the magnetic field $B \rightarrow -B$, the
longitudinal resistances verify the symmetry:\citep{Peled04}
\begin{equation}
\label{e1.3}
R^L_{14,23}(B) = R^L_{14,65}(-B),
\end{equation}
to high accuracy,
although $R^L_{14,23}(B)\ne R^L_{14,23}(-B)$ and $R^L_{14,65}(B)\ne
R^L_{14,65}(-B)$ except in regime A3, on the high-$\nu$ side of the
transition, where the two longitudinal resistances have identical
fluctuation patterns.

In the following, we show that all these observations can be explained
within the non-interacting electron approximation, using a combination
of ideas about the IQHE and mesoscopic transport.

\section{Numerical simulations}
\label{numsim}

\subsection{The conductance matrix and the resistances}

The linear response function of the six-terminal Hall bar is the
$6\times6$ conductance matrix $\hat{g}$ that defines the relationship
between the currents flowing through the various leads, and their
voltages: $I_\alpha =\sum_\beta g_{\alpha \beta} V_\beta$. Here,
$I_\alpha$ is the current in the lead $\alpha=1, \dots, 6$. We use the
convention $I>0$ $ (I<0)$ for currents coming out of (flowing into)
the sample.  $V_\alpha$ is the voltage of the lead $\alpha$.

Knowledge of the conductance matrix allows us to calculate the various
resistances in a straightforward way. In the conventional setup, the
current flows from lead 1 to lead 4 and therefore $\hat{I}_{14} =
\begin{pmatrix} -I & 0 & 0 & I & 0 & 0 \end{pmatrix}^T$.  Without loss
of generality, we set $I=1$ (this is a linear response theory) and
$V_4=0$ (the voltage differences are not affected if one of the
terminals is grounded).  After solving the $6\times6$ equation
$\hat{g} \hat{V} = \hat{I}_{14}$ for the other 5 voltages, we can
directly compute the various resistances from their definitions:
\begin{subequations}
\label{e2.31}
\begin{eqnarray}
\label{e2.31a} R^L_{14,23} = V_2 -V_3, \; R^L_{14,65} = V_6 -V_5, \\
\label{e2.31b} R^H_{14,62} = V_6 -V_2, \; R^H_{14,53} = V_5 -V_3.
\end{eqnarray}
\end{subequations}
The two-terminal resistance is determined similarly. In this case, the
electric currents flow from lead 6 to lead 3: $\hat{I}_{63} =
\begin{pmatrix}0 & 0 & I & 0 & 0 & -I 
\end{pmatrix}^T$.  After solving the equation $\hat{g} \hat{V} =
\hat{I}_{63}$ for the corresponding voltage distribution, we can
calculate immediately the two-terminal resistance (assuming again that
$I = 1$):
\begin{equation}
\label{e2.32}
 R_{2t} = R_{63,63} = V_6 -V_3.
\end{equation}
Clearly, any other measurement can be simulated in a similar fashion
within this formalism. All resistances are functions of various
elements of the conductance matrix.

At $T=0$, the off-diagonal elements of the conductance matrix
$\hat{g}$ are calculated\cite{Baranger89} by solving a multi-channel
scattering problem, according to the Landauer-Buttiker formula:
\begin{equation}
\label{lb}
g_{\alpha,\beta\ne \alpha}(E_F) = \frac{e^2}{h} \sum_{i,j}\left|
t_{\alpha i, \beta j}(E_F)\right|^2.
\end{equation}
Here, $t_{\alpha i, \beta j}(E_F)$ is the amplitude of probability
 (transmission amplitude) that an electron with the Fermi energy
 $E_F$, which is injected into the sample through the $j^{th}$
 transverse channel of lead $\beta$, will scatter out of the sample
 into the $i^{th}$ transverse channel of lead $\alpha$. The sum over
 all the transverse channels of the two leads simply gives the total
 probability $p_{\beta \rightarrow \alpha}(E_F)$ for the electron with
 $E_F$ to scatter from lead $\beta$ into lead $\alpha$, and thus:
%%%%%%%%%%%%%%%%%%%%%%%%%%%%%% EQUATION %%%%%%%%%%%%%%%%%%%%%%%%%%%%%%
\begin{equation}
\label{ne1}
g_{\alpha,\beta\ne \alpha}(E_F) = \frac{e^2}{h} p_{\beta \rightarrow
  \alpha}(E_F).
\end{equation}
%%%%%%%%%%%%%%%%%%%%%%%%%%%%%%%%%%%%%%%%%%%%%%%%%%%%%%%%%%%%%%%%%%%%%%
The diagonal elements of the conductance matrix can then be calculated
using the restrictions imposed by charge conservation and gauge
invariance:\cite{Baranger89} $\sum_\alpha g_{\alpha \beta} =
\sum_\beta g_{\alpha \beta} = 0$.  The diagonal elements are thus
%%%%%%%%%%%%%%%%%%%%%%%%%%%%%% EQUATION %%%%%%%%%%%%%%%%%%%%%%%%%%%%%%
\begin{equation}
\label{diag}
g_{\alpha \alpha} =-\sum_{\beta\ne \alpha}g_{\alpha
\beta}=-\sum_{\beta\ne \alpha}g_{\beta \alpha }
\end{equation}
%%%%%%%%%%%%%%%%%%%%%%%%%%%%%%%%%%%%%%%%%%%%%%%%%%%%%%%%%%%%%%%%%%%%%%

The finite-temperature generalization in the linear regime is straightforward:
\begin{equation}
\label{finT}
g_{\alpha,\beta}(\mu,T) = \int_{-\infty}^{\infty} dE
g_{\alpha,\beta}(E)\left[-\frac{d f(E)}{dE} \right],
\end{equation}
where $f(E) = [\exp(\beta(E-\mu))+1]^{-1}$ is the Fermi distribution,
where $\beta =1/(k_BT)$ and $\mu$ is the chemical potential.

The strategy for the numerical simulations of the IQHE in mesoscopic
samples is thus apparent. We have to solve a complicated,
multi-channel scattering problem to find all the amplitudes of
transmission through the Hall bar. This allows us to calculate the
conductance matrix and therefore the various resistances. This is
repeated for many values of the Fermi energy (corresponding to
different filling factors $\nu$ of the Landau level) to obtain the
the traces of the resistances as $\nu$ changes. We now describe the
details of our simulations.

\subsection{The model}
\label{smodel}

In this section we describe the model we use to numerically simulate
IQHE transitions in mesoscopic samples. This comprises the sample, the
confining potential, the disorder potential, the external leads and
their contacts to the sample. As already stated, we assume
non-interacting electrons -- the usual approximation when treating the
IQHE. For numerical convenience, we ignore Landau level (LL) mixing,
although the whole formalism can be straightforwardly extended to
include it.  LL mixing can be ignored for large magnetic fields (the
case of interest to us), when the cyclotron energy is the largest
energy scale in the problem.  In the following, we take the effective
mass of the electron in the 2DES to be that of GaAs/AlAs
heterostructures, $m^*=0.067m_e$, and use $-e, e>0$ for the electron
charge.

\subsubsection{The sample}

In choosing the sample geometry, we face an apparent problem. On one
hand, when working with Landau levels, it is convenient to consider a
rectangular sample with {\em periodic} boundary conditions along one
axis, since in this case the eigenstates of each LL are known
analytical functions. This simplifies considerably the calculation of various
matrix elements.  On the other hand, however, a Hall bar with
different contacts on all sides, such as sketched in Fig. \ref{fig1},
is obviously not consistent with periodic boundary conditions.

Our solution is to start with an area larger than the Hall bar
itself. For this enlarged area we impose cyclic boundary conditions to
generate a convenient basis for the Hilbert subspace of each LL. We
then add a confining potential to ``carve out'' the Hall bar from this
larger area. This potential is described in the next subsection.

Consider, then, 2D electrons in an area of size $L_x\times L_y$, with
periodic boundary conditions in the $y$-direction, and placed in a
perpendicular magnetic field $B$. The spectrum consists of Landau
levels of energy $E_{n\sigma} = \hbar \omega_c (n+1/2) - g \mu_B B
\sigma$, where $n=0,1,2,...$ is the LL index, and $\sigma=\pm 1/2$ is
the $z$-axis spin projection. The cyclotron energy is $
\hbar\omega_c=eB/m^*$.

If we use the Landau gauge $\vec{A}=(0, Bx, 0)$, the eigenfunctions
for the $(n, \sigma)$ Landau level are:\cite{Zhou04b}
\begin{equation}
\label{2.1aa} \langle \bvec r | n, X, \sigma \rangle = { e^{-i{Xy
\over l^2}} \over \sqrt{L_y}} e^{-{1 \over 2l^2}\left( x -X \right)^2
} { H_n\left({x -X \over l}\right) \over
\sqrt{2^nn!\sqrt{\pi}l}}\chi_\sigma
\end{equation}
where $l = \sqrt{ \hbar c \over eB}$ is the magnetic length, $H_n(x)$
are Hermite polynomials and $\chi_\sigma$ are the $z$-spin
eigenstates.  The cyclic boundary condition in the $y$-direction
require that $X_j = j {2\pi l^2 / L_y } $, $j \in Z$. $X_j$, the
guiding center, characterizes the location at which individual basis
states are centered on the $x$-axis [see Eq.~(\ref{2.1aa})]. Since
$-L_x/2<X_j<L_x/2$, it follows that $-N/2 < j < N/2$, and the
degeneracy of each spin-polarized LL is $N= L_xL_y /(2\pi l^2)$. As we
show later, $N$ roughly defines the size of the matrix involved in
solving the scattering problem (several more states on the contacts
have to be included as well).

In the simulations shown here, we use $L_x=L_y=4\mu$m. For a field of
several Tesla, this leads to a degeneracy $N \sim 10^4$, which can be
easily handled by a generic PC cluster in a reasonable amount of
time. The main problem is that our sample is shorter than the
experimental sample ($4\mu$m instead of $\sim$20$\mu$m). This means
that terminals 1 and 4 (see Fig. \ref{fig1}) are much closer to the
other 4 terminals than in reality. We fix this problem with the aid of
the confining potential, as described in the next subsection.

\subsubsection{The confining potential}

\begin{figure}[t]
\centering \includegraphics[width=0.6\columnwidth,angle=-90]{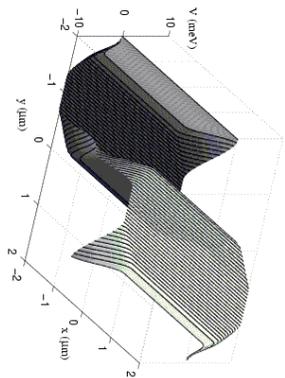}
\caption{The confining potential $V_c(\vec{r})$ that we use. It varies
  smoothly near the $y$-edges, and has triangular barriers in the
  corners. See text for details.}
\label{fig6}
\end{figure}

As stated before, we add a confining potential $V_c(\vec{r})$ to
define the Hall bar from the larger area $L_x \times L_y$ spanned by
the LL Hilbert subspace. We have
tested several functional forms, to see which are physically
reasonable. Some of the choices considered and their pros and cons are
discussed in Appendix \ref{sconf}.

The confining potential used in simulations is shown in
Fig. \ref{fig6}. It is an odd function $V_c(x,y) = -V_c(x,-y)$ and
such that $V_c(x,-L_y/2)=0$. The Hall bar is the region
$-L_y/2<y<0/2$ and $-L_x/2<x<L_x/2$, where
the confining potential is: 
\begin{eqnarray}
  &&\nonumber V_c(x,y )= -V_{\rm gap} { (1-e^{y/\lambda})(1-e^{-{L_y/2+y
    \over \lambda }}) \over 1-e^{-L_y/2 \lambda} } \times \\
    &&\prod_{\nu=\pm1\atop\mu = 0,1}g( |x - \nu L_x/2|,|y + \mu L_y/2|),
\end{eqnarray}
The first line describes the main features: 
the confining
potential is approximatively equal to $-V_{\rm gap}$ inside the Hall
bar. We use $V_{\rm
gap}=10$~meV in our simulations, which is much larger than the
amplitude of the disorder potential (see below). As a result, for
Fermi energies $ E_F <0$, the electrons are confined to the region
$[-L_x/2, L_x/2] \times [-L_y/2,0]$ which defines the Hall bar to
which contacts will be attached. Near the Hall bar $y$-edges at $y=0,-L_y/2$,
the confining potential rises smoothly to zero, on a length-scale
$\lambda$ (chosen to be 40~nm in our simulations; see  discussion in Appendix
\ref{sconf}). The second line describes the triangular potential
barriers added in the four corners of the Hall bar:
\begin{eqnarray}
\nonumber g(x,y) &=& \sin^2 [(x/l_x + y/l_y)\pi/2]\Theta(1-x/l_x -
y/l_y) + \\ &&\Theta(x/l_x+ y/l_y-1),
\end{eqnarray}
where $l_x$ and $l_y = L_y/6$  define their spatial extension.
These barriers help isolate the 
contacts 1 and 4 from the other contacts, and thus compensate for the
shortness of our sample. The alternative is to use a longer sample;
this, however,  requires significantly more CPU time.

The symmetric choice we made for the confining potential is not
necessary; however, it is advantageous for several reasons. If we
take the disorder potential such that $V_d(x,-y)=-V_d(x,y)$ as well,
then the particle-hole symmetry of the total potential for the whole
sample means that all quantities must be symmetric with respect to
$E_F=0$. This allows us to check and calibrate the numerical
procedure, as discussed in Appendix \ref{sconf}.

Another advantage of a symmetric choice is that it allows us to easily
define the Hall bar filling factor for $E_F<0$, as being twice the
filling factor of the LL (same number of filled states, but half the
area). Strictly speaking, this is an underestimate. At the Fermi
energies $E_F \sim -V_{\rm gap}$ of interest to us (see below), the
effective area of the Hall bar is somewhat smaller, because of the
shape of the confining potential.

\subsubsection{The disorder potential}

\begin{figure}[t]
\includegraphics[width=0.6\columnwidth,angle=-90]{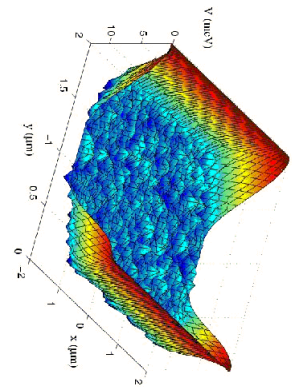}
\caption{(color online) Total confining plus disorder potential inside the
   Hall bar, $-L_y/2<y<0$ (compare with Fig.~\ref{fig6}).}
\label{fig8}
\end{figure}

The experimental sample contains many In impurities which serve as
short range scattering centers. We simulate a short-range disorder
potential $V_d(\vec{r})$ by adding 18000 Gaussian scatterers randomly
placed inside the Hall bar area $[-2,2] {\rm \mu m} \times [-2,0]{\rm
\mu m}$. Each Gaussian scattering potential is of the form
$Ae^{-r^2/d^2}$, where $d$ is uniformly distributed in the interval
$[0.01,~0.03)~ \mu$m, and $A$ is uniformly distributed between $[-0.6,
0.6]$~meV.

A typical disorder realization within the Hall bar is shown in
Fig. \ref{fig8}, added to the confining potential for the Hall bar side of
the total sample. The disorder is very rough and short-range, almost
white-noise-like. Moreover, its amplitude is much smaller than that of
the confining potential, as required in order to have a well defined 
Hall bar.  Our simulations scan a dense energy grid, typically in the
range $[-10.3, -9.7]$~meV, close to a filling factor $\nu \sim 1/2$
for the Hall bar.

\subsubsection{The leads and the contacts}

The IQHE is independent of the type of current-carrying external leads
used and of their contacts to the sample, provided that (i) the leads
are reasonably good metals and (ii) the contacts allow easy transfer
of the electrons into and out of the Hall bar. As a result, no
detailed realistic modeling of the leads and of the contacts is
required.  Instead, we can use simple, idealized models that satisfy
requirements (i) and (ii).

As in Ref.~\onlinecite{Zhou04b}, we model each external lead as a
collection of independent, semi-infinite, perfectly metallic
one-dimensional tight-binding chains.  Each such chain represents a
transverse channel of that lead, which carries currents independently
of the other channels in the lead. The choice for the hopping matrix
$t$ and on-site energy $\epsilon_0$ that define the Hamiltonian of
each such chain, has been discussed in detail in
Ref.~\onlinecite{Zhou04b}. Briefly, $t$ must be chosen so as to
minimize the contact resistance. We use $t=0.1$~meV,  comparable to
the major matrix elements of the Hamiltonian matrix 
elements inside the sample.  To insure that the leads can always carry
currents to and from the Hall bar, we use a ``floating'' spectrum,
i.e. we set $\epsilon_0=E_F$ that we investigate. With these choices,
requirement (i) is trivially satisfied.

The contacts are represented by matrix elements connecting states on
the leads (tight-binding sites) to {\em contact states} in the Hall
bar, which are appropriate linear combinations of the LL basis
states. The simplest choice is to add to the total Hamiltonian a
hopping term of the form $ -t (c_0^\dagger d +h.c. )$ for each chain
(transverse channel). Here, $c_0^\dagger$ is the creation operator for
the last site of the chain, while $d$ is the annihilation operator
for its corresponding contact state.

The contact states are chosen so as to satisfy the following general
requirements:

(1) Each contact state must be localized close to the region of the Hall
bar where the corresponding lead is connected.

(2) Different leads have orthogonal contact states, but channels of
the same lead may have non-orthogonal contact states. This is because
the direct connection (short circuit) between different leads is
forbidden.

(3) For numerical efficiency, it is convenient to preserve the
sparsity of the total Hamiltonian. We therefore require that a contact
state is a linear combination of a small subset of the $N$ LL basis
states.

For leads 1 and 4 (the current source and drain, see Fig. \ref{fig1}),
the contact states must be localized close to the $x=\pm L_x/2$ ends
of the sample. The LL basis states $|n, X_j, \sigma\rangle$ with $j
\sim \pm N/2$ are exactly such states. For simplicity, we take each of
them to be a contact state, as in
Ref.~\onlinecite{Zhou04b}. Specifically, we assume that leads 1 and
4 have $N_c$ transverse channels each (we use $N_c=100$ in the
simulations shown here). The $m^{\rm th}$ transverse channel of the
lead 1 (4) is coupled through a hopping term to the LL basis state
$|n, X_j, \sigma\rangle$ with $j = -N/2+m$, and respectively $j =
N/2-m$, $ (\forall)~1\le m \le N_c$. With this choice, all contact
states are orthogonal to one another, and all three requirements
listed above are satisfied.  This means that electrons can be
injected/removed from the Hall bar anywhere within a strip of width
$N_c 2\pi l^2/L_y$ from the $x=\pm L_x/2$ edges, through contacts 1
and 4.

\begin{figure}[t]
\centering \includegraphics[width=0.6\columnwidth,angle=-90]{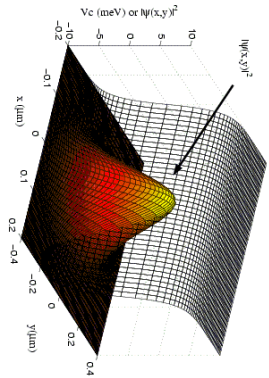}
\caption{(color online) Schematic drawing of the probability density of a contact
state  on the $y$-edge. It overlaps with the entire region where the
confining potential rises from $-V_{\rm gap}$ to $V_{\rm gap}$, so that it can
inject/collect electrons of any Fermi energy from the Hall bar.}
\label{fig7}
\end{figure}

For the voltage probes (leads 2, 3, 5 and 6) we need a different
strategy, since they are located on the $y=-L_y/2$, respectively $y=0$
edges of the Hall bar. The contact states are chosen to be of the
form:
\begin{equation}
 d^\dag |{\rm vacuum}\rangle = \lambda \sum_m e^{-{ (X_m-x_c)^2 \over
a^2} + im \theta} \ket{n, X_m, \sigma},
\label{e2.2}
\end{equation}  
where $a$ is a parameter that controls the $x$ and $y$ extension of
the wave-function, $\theta = 0, \pi$ for $y=0, -L_y/2$ respectively,
and $\lambda$ is the normalization factor. As $a$ changes, the shape
of this wavefunction changes from a strip in the $x$-direction
to a strip in the $y$-direction with fixed total area. We choose $a = 0.2
l$   so that the probability distribution of the
contact states overlaps with the region of smooth variation  of the confining
potential near the $y$-edges. In this case, the profile of the contact
states is  an ellipse with its
longer axis in the $y$-direction (see Fig.~\ref{fig7}).

In Eq.~(\ref{e2.2}), $x_c$ specifies the location of the contact state
along the $y$-edge. We use $N_c/4$ ($25$, in these simulations)
tight-binding chains for each of 
these leads, and correspondingly $N_c/4$ different contact states.  For
leads 3 and 5, the centers $x_c$ of their contact states are
distributed uniformly in the range  $[1, 1.5]\mu$m, while for leads 2
and 6 the 
range is $[-1.5, -1]\mu$m. The summation in Eq.~(\ref{e2.2}) is
truncated so that no pair of neighboring contact states contain the
same LL basis states in their expansions. The contact states on the $y =
-L_y/2$ edge (for leads 2 and 3), are magnetic translations of the
corresponding 
states of leads 5 and 6. The alternating phase factor $e^{im
\pi}$ insures that  contact states on different $y$-edges  are
orthogonal to one another, so that short-circuits are avoided.

\subsection{Solving the scattering problem}

The total Hamiltonian is the sum of the sample Hamiltonian, the
Hamiltonians for the 6 leads, and the terms describing the contacts
between sample and leads:
%%%%%%%%%%%%%%%%%%%%%%%%%%%%%% EQUATION %%%%%%%%%%%%%%%%%%%%%%%%%%%%%%
\begin{equation}
\label{ne2}
{\cal H}={\cal H}_{\rm sample} + \sum_{\alpha=1}^{6}{\cal
  H}^{(\alpha)}_{\rm lead} +{\cal H}_{\rm contacts}
\end{equation}
%%%%%%%%%%%%%%%%%%%%%%%%%%%%%%%%%%%%%%%%%%%%%%%%%%%%%%%%%%%%%%%%%%%%%%
The numerical results presented here are obtained in the lowest Landau
level (LLL) with $n=0$ (the spin projection is
irrelevant. Calculations in higher LL proceed in a similar
fashion). For simplicity, in the following we denote the LLL basis
states by $|X_j\rangle = c_j^\dagger|{\rm vacuum}\rangle \equiv| n=0,
X_j, \sigma\rangle$.  In the Hilbert subspace of the spin-polarized
LLL, the Hamiltonian of the sample is:
\begin{equation}
\label{e2.3} 
{\cal H}_{\rm sample} = \sum_{i=-N/2}^{N/2}\sum_{j=-N/2}^{N/2} \langle
X_i |V_c + V_d | X_j \rangle \cdot c^\dagger_ic_j\;,
\end{equation}
where $N = L_xL_y /(2\pi l^2)$ is the number of states in a LL and an
overall constant LLL energy shift $E_{0\sigma}$ has been ignored.  The
matrix elements $\langle X_i |V_c + V_d | X_j \rangle$ of the
confining and disorder potentials are computed as described in
Ref.~\onlinecite{Zhou04b}.  Briefly, the idea is that matrix
elements of a plane-wave are simple analytical functions:
$$ \bra{ X_i} e^{i\bvec{q\cdot r}}\ket{X_j} = \delta_{X_i,X_j-q_yl^2}
e^{{i\over 2}q_x(X_i+X_j)}e^{-{1 \over 2} Q}
$$ where $Q = {1 \over 2}l^2(q_x^2+q_y^2)$. (The generalization for
higher LLs and/or LL mixing is straightforward, see for instance
Eq.~(7) in Ref.~\onlinecite{Zhou04b}). The strategy, then, is to
perform a fast Fourier transform of the potentials on a grid dense
enough to reproduce them with high accuracy while maintaining the
sparseness of the Hamiltonian. The matrix element of each Fourier
component is given by the previous formula, allowing one to
efficiently compute the matrix elements of the confining and
disorder potentials. Note that for higher LLs, the matrix elements
differ only by a Laguerre polynomial. This is not likely to influence
the physics in a significant way and suggests that  numerical results for
higher LLs should be qualitatively similar to those we present here for the
LLL.

The leads are collections of $N_c^\alpha$ semi-infinite tight-binding
chains, each describing a transverse channel:
%%%%%%%%%%%%%%%%%%%%%%%%%%%%%% EQUATION %%%%%%%%%%%%%%%%%%%%%%%%%%%%%%
\begin{equation}
\label{ne3}
{\cal H}^{(\alpha)}_{\rm lead} = \sum_{i=1}^{N_c^{\alpha}}
 \sum_{m=0}^{\infty} \left[\epsilon_0 c^{(\alpha) \dagger}_{i,m}
 c^{(\alpha)}_{i,m} - t \left(c^{(\alpha) \dagger}_{i,m}
 c^{(\alpha)}_{i,m+1}+h.c.\right) \right]
\end{equation}
%%%%%%%%%%%%%%%%%%%%%%%%%%%%%%%%%%%%%%%%%%%%%%%%%%%%%%%%%%%%%%%%%%%%%%
Thus, $c^{(\alpha) \dagger}_{i,m}$ creates an electron on the site $m
\ge 0$ of the $i^{\rm th}$ transverse channel of lead $\alpha$. We use
the convention that site $m=0$ is always the end site of the chain.
The solution we present can be trivially generalized to allow for
different $\epsilon_0$ and $t$ parameters for each channel, as well as
longer-range hopping. However, since the physics in the sample is
independent of the lead details (as long as the leads can carry
currents to and from the sample) the simple choice we make should be
sufficient. The choices for $t, \epsilon_0$ and number of channels
$N_c^\alpha$ for each lead have been discussed  previously.

Finally, the contacts are described by matrix elements between sites
$m=0$ of the various channels and their corresponding contact states:
%%%%%%%%%%%%%%%%%%%%%%%%%%%%%% EQUATION %%%%%%%%%%%%%%%%%%%%%%%%%%%%%%
\begin{equation}
\label{ne5}
{\cal H}_{\rm contacts} = -t
 \sum_{\alpha=1}^{6}\sum_{i=1}^{N_c^{\alpha}}
 \left[c^{(\alpha)\dagger}_{i,0}d^{(\alpha)}_{i} +h.c. \right]
\end{equation}
%%%%%%%%%%%%%%%%%%%%%%%%%%%%%%%%%%%%%%%%%%%%%%%%%%%%%%%%%%%%%%%%%%%%%%
where, as described in the previous section, each contact state is a linear
combination of LLL basis states, which is localized in the appropriate
region of the sample. Again, generalizations for more complex contacts
can be easily incorporated, but should not be necessary.

We now have to solve a scattering problem for an electron injected
with energy $E$ in the channel $j$ of lead $\beta$, which means that
we must find an eigenstate  ${\cal H}|\Psi_{\beta j}\rangle =
E|\Psi_{\beta j}\rangle$ of the form
$$ |\Psi_{\beta j}\rangle =
\left[\sum_{\alpha=1}^{6}\sum_{i=1}^{N_c^\alpha} \sum_{m=0}^{\infty}
  \phi^{(\alpha)}_{i,m} c^{(\alpha) \dagger}_{i,m} +
  \sum_{j=-N/2}^{N/2} \phi_j c^\dagger_j \right]|{\rm vacuum}\rangle
$$ where -- with our convention for indexing channel sites --
$\phi^{(\alpha)}_{i,m} = t_{\alpha i, \beta j}e^{i k m} + \delta_{ij}
\delta_{\alpha\beta} e^{-i km }$. In other words, there is an
in-coming wave with unit amplitude in channel $(\beta j)$, and
out-going waves with various transmission amplitudes in all other
channels. The momentum $k >0$ is such that $E = \epsilon_0 - 2t \cos
k$, i.e. it is the momentum of an electron with energy $E$ propagating
on any of the tight-binding chains.

We solve this scattering problem by recasting it as a {\em
finite-size, inhomogeneous} system of linear equations, which can be
easily solved numerically. The unknowns are the amplitudes $\phi_j$ to
find the electron in various LL states in the sample, plus 5 terms
$\phi^{(\alpha)}_{i,m}$, $m=0,\dots, 4$ for each channel (clearly, if
one knows the wave-function at the first few sites of a tight-binding
chain, one 
knows the wave-function along the entire chain).  This solution is
explained in detail in Ref.~\onlinecite{Zhou04b}, for two leads with
multiple channels. The redistribution of the channels amongst more
leads is trivial, since it only involves changing some of the matrix
elements to the appropriate contact states.

The solution of this inhomogeneous system of linear equations, then,
directly gives us the transmission coefficients $t_{\alpha i, \beta
j}(E)$ from which we compute the conductance matrix [see
Eq.~(\ref{lb})]. This is repeated for many Fermi energies $E_F$
corresponding to various filling factors $\nu$, to obtain the
dependence $\hat{g}(\nu)$ of the various elements of the conductance
matrix on the filling factor.

\section{Numerical Results}
\label{numres}
\subsection{The conductance matrix for the LLL}

\begin{figure}[t]
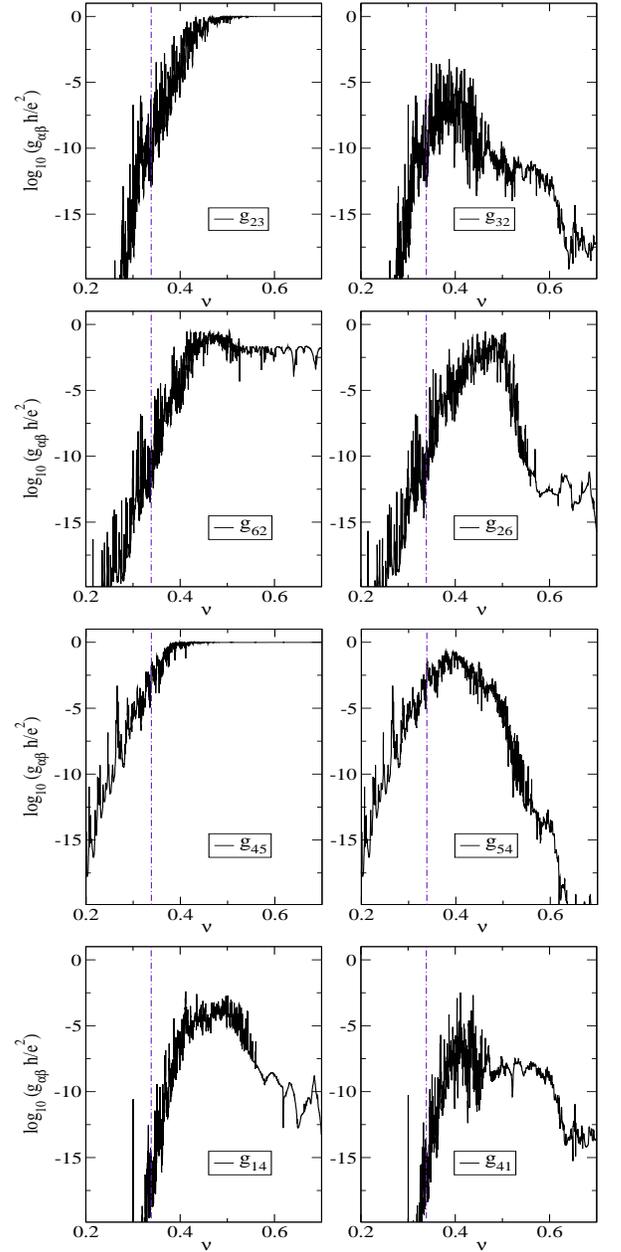

  \centering \includegraphics[width=0.9\columnwidth,
  height=1.6in]{FIG5a.eps}
  \vspace{0.05in} \centering \includegraphics[width=0.9\columnwidth,
  height=1.6in]{FIG5b.eps}
   \vspace{0.05in} \centering \includegraphics[width=0.9\columnwidth,
  height=1.6in]{FIG5c.eps}
   \vspace{0.05in} \centering \includegraphics[width=0.9\columnwidth,
  height=1.6in]{FIG5d.eps}
  \caption{Representative conductance matrix elements, in units of
  $e^2/h$, as a function of  $\nu$. The left panels
  show $g_{23}$, $g_{62}$, $g_{45}$, $g_{14}$, while  the right panels show
  $g_{32}$, $g_{26}$, $g_{54}$, $g_{41}$.  The top
  three left (right) panels characterize transport in  (against)
  the direction of the edge currents. The bottom panels show the
  negligible direct transport between the current source and drain.
  For each pair, the traces are
  almost identical on the left, but 
  very different on the right of the vertical line.}
  \label{fig15}
\end{figure}

Figure \ref{fig15} shows numerical results for various matrix elements
${g}_{\alpha\beta}(\nu)$ inside the LLL $(0 < \nu < 1)$. The
calculation is for the disorder potential shown in Fig. \ref{fig8},
for a magnetic field $B=3$T. 
Two facts are immediately apparent. First, for low filling factors ($\nu
\le 0.34$ for this disorder realization), the conductance matrix is
symmetric to 
high accuracy, 
${g}_{\alpha\beta} = {g}_{\beta\alpha}$. From Eq.~(\ref{ne1}) it
follows that $p_{\beta \rightarrow \alpha}= p_{ \alpha \rightarrow
\beta}$, i.e. the electron has the same (very small) probability to
scatter in both directions. This is expected, since we know that at
low filling factors, the states in the LLL are localized. Scattering
from one terminal to another is only possible through tunneling of the
electron from a state localized near one terminal, into a state
localized near the other terminal. Such tunneling amplitudes are very
small, explaining the small total probabilities for such
scattering. Tunneling is also time-reversal symmetric -- it happens
with  the
same probability  in both directions, consistent with the
symmetry of the conductance matrix. These ideas will be made more
precise in Section \ref{sglparam}, where we analyze the general
structure allowed for the conductance matrix.

The second observation is that at high-filling factors $\nu
\rightarrow 1$, we have $g_{\alpha, \alpha+1} \rightarrow e^2/h$
(where if 
$\alpha=6, \alpha+1 =1$) while all other off-diagonal matrix elements
become vanishingly small. In other words, an electron injected through
terminal $\alpha+1$ scatters with unit probability into terminal
$\alpha$. This behavior clearly signals the appearance of the {\em edge
states} at higher filling factors. These are chiral states,
carrying electrons only in the direction consistent with the
orientation (sign of) the magnetic field $B$. These states are
localized on the edges of the sample, on the ``vertical walls'' created by the
confining potential.

Using Eq.~(\ref{diag}), this shows that in the limit $\nu \rightarrow 1$,
the conductance matrix converges to the simple form:
\begin{equation}
\label{e2.34}
\hat{g}(\nu)\stackrel{\nu \rightarrow 1}{\longrightarrow}
  \hat{g}^{(0)} = {e^2 \over h} \begin{pmatrix} -1 & 1 & 0 & 0 & 0 & 0
  \\ 0 & -1 & 1 & 0 & 0 & 0 \\ 0 & 0 & -1 & 1 & 0 & 0 \\ 0 & 0 & 0 &
  -1 & 1 & 0 \\ 0 & 0 & 0 & 0 & -1 & 1 \\ 1 & 0 & 0 & 0 & 0 & -1
\end{pmatrix}.
\end{equation}
The resistances corresponding to this asymptotic limit are
straightforward to find. Solving $\hat{I}_{14} = \hat{g}^{(0)}\cdot
\hat{V}$ for $I=1, V_4=0$, we find $V_5 = V_6= h/e^2$, $V_2 = V_3 =
0$. Eqs. (\ref{e2.31a}), (\ref{e2.31b}) then give $ R^H_{14,62} =
R^H_{14,53} = h/e^2$, $R^L_{14,23} = R^L_{14,65} = 0$. In other words,
the values expected for the first IQHE plateau, which indeed is established
when the LLL is more than half-filled.

\subsection{Fluctuations of resistances in the  LLL}

\begin{figure}[t]
  \centering \includegraphics[width=\columnwidth]{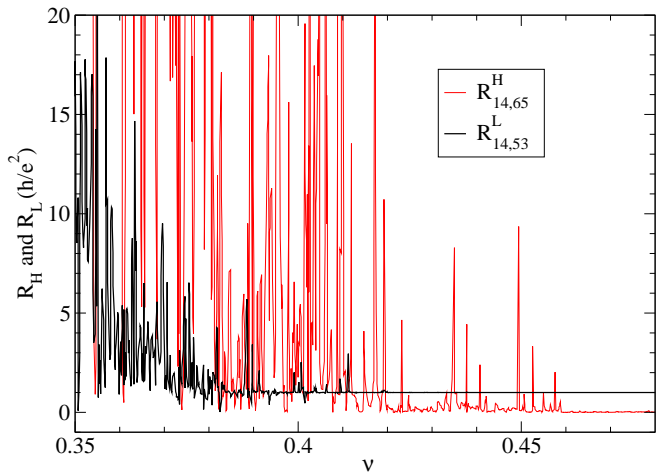}
  \caption{A pair of $R^L$ and $R^H$ for the disorder shown in
    Fig. (\ref{fig8} and 
    $0<\nu<1$. $R^L$ shows strong fluctuation before quantization of
    $R^H$ is destroyed.  The data are calculated at $T=0$.}
\label{fig13}
\end{figure}

\begin{figure}[t]
  \centering \includegraphics[width=\columnwidth]{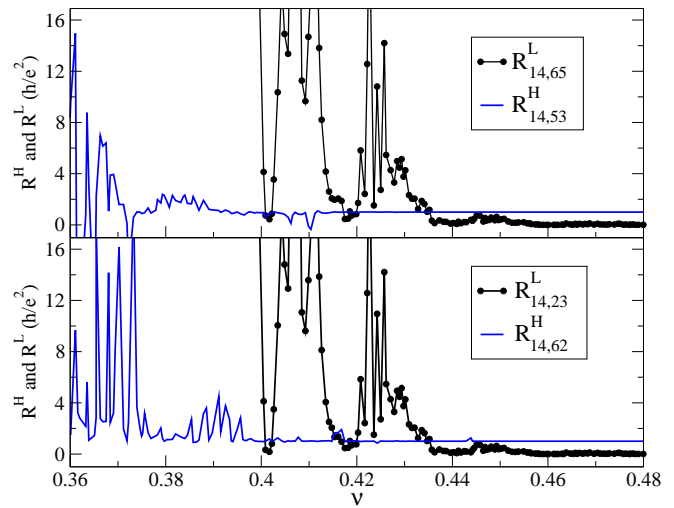}
  \caption{Both pairs of $R^L$ and $R^H$ for a different disorder.
    Both $R^L$ show strong fluctuations before the quantization of the
    $R^H$ disappears. The results correspond to $T =11.6$mK.}
\label{fig14}
\end{figure}

The fluctuations of the various resistances are due to variations of
$\hat{g}(\nu)$ from the asymptotic limit $\hat{g}^{(0)}$. Using
$\hat{g}(\nu)$ shown in Fig. \ref{fig15}, we calculate the resistances
for all filling factors $0 < \nu < 1$. In Fig.~\ref{fig13} we plot the
pair $R^H_{14,65}$ and $R^L_{14,53}$ as a function of $\nu$ near
half-filling.    Three different regimes are apparent: for
$\nu>0.46$, $R^H=h/e^2$ and $R^L=0$, corresponding to the first IQHE
plateau. For $0.42<\nu<0.46$, $R^L$ exhibits large fluctuations,
however $R^H$ is still well quantized. This is precisely the type of
behavior observed experimentally\cite{Peled03a} (see Section
\ref{sec2}, point B). For $\nu < 0.42$ the transition to the
insulating phase occurs, and all resistances increase sharply as the
wave-functions at the bottom of the LLL become more and more localized,
leading to progressively smaller values for the matrix elements of
$\hat{g}$ (see Fig. \ref{fig15}). This results in large fluctuations
of both resistances at $T=0$; these are smeared out at finite $T$,
as shown in Fig.~\ref{fig14}.

Figure~\ref{fig14} shows all four resistances, for a different
disorder realization, at $T=11.6$mK.  The two Hall resistances are
quantized to $h/e^2$ down to $\nu\approx 0.4$.  Below $\nu \approx
0.46$, both $R^L$ exhibit fluctuations which quickly evolve into high
peaks around $\nu = 0.42$ where both $R^H$ still have only minor
deviations from $h/e^2$. Comparing the upper and lower panel of
Fig.~\ref{fig14}, we see that the fluctuations of $R^L_{14,23}$ and
$R^L_{14,65}$ are almost identical in this regime. Below $\nu\approx
0.4$ is the transition to the insulating phase.

The resistances in the insulating phase in Figs.~\ref{fig13} and
\ref{fig14} are not perfectly similar to the experimental
results.\cite{Peled03a} In the experiment, even at the lowest
temperature, the rise in $R^L$ is less steep and less noisy than the
behavior shown by our simulations. This is because when the Fermi
energy resides amongst the low-$\nu$, strongly localized states of the LLL, the
charge transport described here is no longer the dominant mechanism
for conduction. Temperature assisted hopping conductivity and other
inelastic scattering processes are likely to contribute to the
conductance matrix more than the ``band'' conductance matrix which the
simulation actually calculates. Therefore, we do not claim to have a realistic
description of the insulating phase here. However, closer to
half-filling the matrix elements of $\hat{g}$ become reasonably
large, elastic scattering becomes the dominant mechanics for
conduction and  our simulations  describe well  the behavior observed
experimentally.

\subsection{The conductance matrix of higher LL}

The conductance matrix when the Fermi level is in a higher LL can be
obtained from the conductance matrix of the LLL using the
superposition principle.\cite{Tsui97} We simply add a contribution
$\hat{g}^{(0)}$ for each completely filled LL to the matrix
$\hat{g}(\nu)$ describing the response of the LL hosting the Fermi
energy (in this context, $0<\nu<1$ denotes the filling factor of the LL hosting
the Fermi energy, not the total filling factor). As discussed,
$\hat{g}^{(0)}$ describes transport 
through the 
edge states, which is the only possible contribution of a completely
filled LL. Indeed, we see that if $E_F$ is such that $n$ LL are
completely filled, according to the superposition principle we have
$\hat{g} = n\hat{g}^{(0)}$. This immediately leads to the solutions
$R^H =h/(ne^2)$, $R^L =0$, i.e. the $n^{\rm th}$ IQHE plateau, as
expected.

The validity of the superposition principle (which we take for granted
here), is based on the belief that the IQHE transition can be
described within one single LL, and that each plateau-to-plateau
transition is in the same universality class. Wei {\em et al.}\cite{Tsui88}
experimentally confirmed this hypothesis. Shahar {\em et al.}\cite{Tsui97}
further developed this idea by mapping the transition between adjacent
IQHE plateaus to the insulator-to-QH transition in the LLL.

The superposition principle relies on several conditions, which are
satisfied in our simulations: (1) One must confirm that for a single
LL, as the Fermi energy is raised and $\nu \rightarrow 1$, the
conductance matrix $\hat{g} \rightarrow \hat{g}^{(0)}$ at reasonable
filling factors, close to the center of the LL. We have already shown
that our simulations satisfy this for the LLL. Higher LL should behave
similarly, since the only difference is a simple change in the matrix
elements of the disorder plus confining potential [see
Eq.~(\ref{e2.3}) and discussion following it].  (2) The cyclotron gap
is large enough to justify neglect of the LL mixing. This is certainly
valid for $\nu \sim n + 0.5$, because here the charge transport is
dominated by processes in the bulk of the Hall bar, where the disorder
is much smaller than the cyclotron gap. However, for the spatially
close edge-states of different LLs, one does expect inter-Landau level
and spin-dependent scattering to become
relevant.\cite{note} As a result, we
do not claim that the edge-states produced in the single LL simulation
are necessarily realistic.  However, this has no influence on the
conductance matrix. The reason is simple: there is no backscattering
among the edge-states.\citep{Buttiker88} All edge-states near the same
boundary of the sample carry currents in the same direction,
irrespective of their LL index and spin-polarization. Scattering among
these edge-states results in a unitary transformation which conserves
the total probability to carry electrons from one terminal to the
next. As a result, even though inclusion of the filled LL states in the
simulation may change some details of the edge-state wave-functions,
the total conduction matrix $\hat{g}$ is not affected. This, in conjunction
with the significant economy of CPU time when LL mixing is neglected,
explains why we perform the simulations for a single LL.

\subsection{Fluctuations of resistances in higher LL}

Using the superposition principle, the fluctuations of the resistances
for $1 < \nu < 2$ can be calculated using the conductance matrix
$\hat{g}(\nu-1) + \hat{g}^{(0)}$, where $\hat{g}(\nu)$ is the LLL
conductance matrix already analyzed  and the second term is the
contribution of the  filled LL.

\begin{figure}[t]
  \centering \includegraphics[width=0.95\columnwidth]{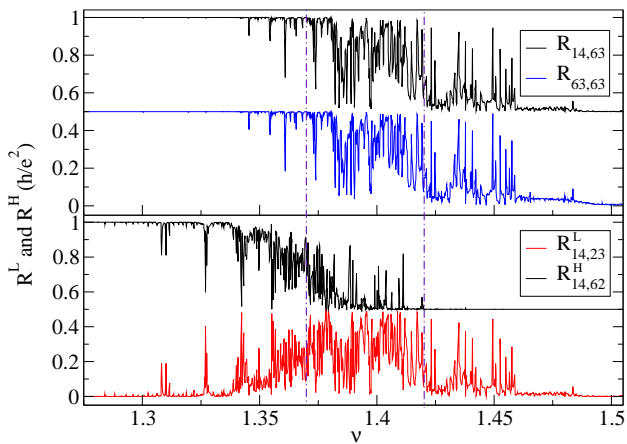}
  \caption{A typical transition from the first to the second IQHE
  plateau at $T=0$. Lower panel: $R^L_{14,62}$ and $R^H_{14,23}$ in units of
  $h/e^2$. Upper panel: The sum $R^L_{14,62}+R^H_{14,23}=R_{14,63}$ of
  the resistances shown in the 
  lower panel, and $R_{2t}= R_{63,63}$ (shifted by $-0.5{h/e^2}$).
  Vertical lines indicate the 
  boundaries of the critical region. See text for more details. }
  \label{fig11}
\end{figure}

The resistances $R^H_{14,62}$ and $R^L_{14,23}$ corresponding to the
disorder shown in Fig. \ref{fig8} are plotted in the lower panel of
Fig. \ref{fig11}. The three regimes found experimentally (see
Sec.~\ref{sec2}, points A1 -- A3) are clearly observed within the
transition from the first to the second IQHE plateau.  Their
(approximate) boundaries are marked by vertical lines in
Fig.~\ref{fig11}, as a guide to the eye.  At low-$\nu$ (high-$B$), the
fluctuations of $R^H$ and $R^L$ are correlated such that $R^L+R^H=h/e^2$. This
is seen more clearly in the upper panel, where their sum
$R^H_{14,62}+R^L_{14,23}=R_{14,63}$ is plotted.  At high-$\nu$
(low-$B$) $R^H={h/( 2e^2)}$ is quantized while $R^L$ exhibits strong
fluctuations. In the intermediate regime, both $R^H$ and $R^L$ have
strong, uncorrelated fluctuations. The upper panel of
Fig.~\ref{fig11} compares $R^L+R^H $ with
$R_{2t}$ ($R_{2t}$ is shifted by $0.5h/e^2$).  As found
experimentally,\citep{Peled03b} (Sec.~\ref{sec2}, point C) the two
curves are almost identical. Only minor differences at high-$\nu$ are
visible.

\begin{figure}[t]
  \centering \includegraphics[width=\columnwidth]{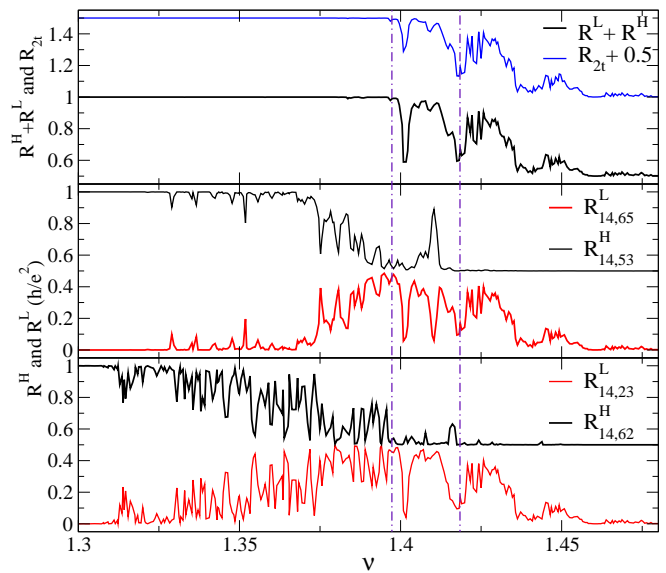}
  \caption{The result of a simulation performed at $T =11.6$mK.  (a)
    Comparison between $R_{2t}$ and $R^H+R^L$. Here, $R_{2t}$ has been
    shifted up for clarity. (b), (c) the two pairs of $R^L$ and $R^H$
    for $1<\nu<2$. The vertical lines indicate the boundary of three
    distinctive fluctuation regimes.}
  \label{fig12}
\end{figure}

Figure~\ref{fig12} shows a set of data calculated for $T = 11.6$~mK
with a different disorder realization. Both pairs of resistances are
shown (the lower panels).  The thermal average was performed using
Eq.~(\ref{finT}).  At finite $T$, the traces are much smoother than at
$T=0$.  The scale of the fluctuations in Fig.~\ref{fig12} is
comparable to that of the experimental data in
Ref.~\onlinecite{Peled03b} -- an indication that our simulations
provide a good approximation of the experimental conditions. While the
fluctuation patterns are clearly sample (disorder) specific, the three
distinct fluctuation regimes are again clearly seen. The comparison
between $R_{2t}$ and $R_{14,63}$ (upper panel)  shows that they are
approximately equal over the entire transition.  Moreover, it is also
apparent that the two pairs of $R^H$ and $R^L$ have different
fluctuation patterns, except on the high-$\nu$ side where the two
$R^L$ are almost identical. Another observation in all our simulations
is that everywhere during this $1< \nu<2$ transition, $0 \le R^L \le
0.5h/e^2$ and $0.5h/e^2< R^H <h/e^2$. Within experimental error bars,
the experimental resistance traces also stay within these
limits.\cite{Peled03b} Thus, all observations are in agreement with
the experimental facts.

\section{Generalized Landauer-B\"uttiker model}
\label{sglparam}

So far, we have shown that our numerical simulations faithfully
reproduce the experimental results. In this section we explain the
physics underlying these various types of correlations, using a simple
but very general model. The consequences of changing the orientation
of the magnetic field, $B\rightarrow -B$, are also addressed.

Since all resistances are functions of the various conductance matrix
elements, it is clear that the correlations of the various resistances are
encoded in the structure of the conductance matrix $\hat{g}$. Thus, it
is important to understand  the allowed structure of the
conduction matrix, based on the various constraints it must satisfy,
and the physics known to be relevant for transport in a LL.

As already discussed, the conductance matrix in the presence of $n$
 filled LLs is of the form $n\hat{g}^{(0)} + \hat{g}(\nu-n)$,
where the first part is the edge-state contribution of the filled LL
levels, and the second term is the contribution of the LL hosting the
Fermi energy. The challenge, then, is to understand
$\hat{g}(\nu)$. Its matrix elements satisfy the following general
constraints: (i) since they are proportional to transmission
probabilities, all off-diagonal matrix elements are positive numbers
$g_{\alpha, \beta\ne \alpha} \ge 0$ [see Eq.~(\ref{ne1})]; (ii) Let
$N_T\ge 2 $ be the number of terminals ($N_T=6$ is the case of
interest to us). The $N_T(N_T-1)$ off-diagonal matrix elements must
satisfy the $2N_T$ constraints of Eq.~(\ref{diag}).  Half of them fix
the value of the diagonal matrix elements, leaving a total of $N_T-1$
supplementary constraints for the off-diagonal elements (one
constraint is trivially satisfied if the other $2N_T-1$ hold).

If $N_T=2$, this implies immediately that the conductance matrix must
be {\em symmetric}, i.e. the most general possible form is:
%%%%%%%%%%%%%%%%%%%%%%%%%%%%%% EQUATION %%%%%%%%%%%%%%%%%%%%%%%%%%%%%%
\begin{equation}
\label{m1}
\hat{g}_{2} = G \left(
\begin{array}[c]{cc}
 -1& 1\\ 1 & -1\\
\end{array}
\right)
\end{equation}
%%%%%%%%%%%%%%%%%%%%%%%%%%%%%%%%%%%%%%%%%%%%%%%%%%%%%%%%%%%%%%%%%%%%%%
 $G$ is the total conductance between the two terminals, since in this
 geometry we must have $I_1=-I_2 = -I$ (all the current injected
 through one terminal must be removed through the other terminal), and
 therefore
$$ \left(
\begin{array}[c]{c}
I_1 \\ I_2 \\
\end{array}
\right) =\hat{g}_{2} \cdot \left(
\begin{array}[c]{c}
V_1 \\ V_2 \\
\end{array}
\right)\rightarrow I= G(V_1-V_2)
$$

For $N_T=3$, the situation is more interesting. First, we separate the
``symmetric'' contributions. Let us denote
$$ G_{\alpha\beta} = \min\left(g_{\alpha\beta}, g_{\beta\alpha}
\right)
$$ We can then rewrite:
$$ \hat{g}_3 = G_{12} \left(
\begin{array}[c]{ccc}
-1 & 1 & 0\\ 1& -1& 0\\ 0 &0 & 0 \\
\end{array}
\right) + G_{13} \left(
\begin{array}[c]{ccc}
-1 & 0 & 1\\ 0& 0& 0\\ 1 &0 & -1 \\
\end{array}
\right)
$$
$$ + G_{23} \left(
\begin{array}[c]{ccc}
0 & 0 & 0\\ 0& -1& 1\\ 0 &1 & -1 \\
\end{array}
\right) + \Delta \hat{g}_3
$$ If the system is time-reversal symmetric, then $\Delta \hat{g}_3=0$
 since transport must proceed with equal probability in any two
 opposite directions, $p_{\alpha \rightarrow \beta} = p_{\beta
 \rightarrow \alpha}$. However, in the presence of a magnetic field,
 there is a preferred direction of charge transport chosen by the
 magnetic field. Thus, in cases of interest to us the conductance
 matrix is generally not symmetric (as already exemplified by
 $\hat{g}^{(0)}$).

The matrix $ \Delta \hat{g}_3$ which contains the non-symmetric
contributions must still satisfy the constraints of Eq.~(\ref{diag}),
since the symmetric contributions do. Moreover, by construction,
exactly 3 of its 6 off-diagonal matrix elements are zero. The
remaining (positive) matrix elements must be all equal, and such that
there is one on each row and on each column, otherwise the constraints
cannot be satisfied. The only possibilities are either
$$ \Delta \hat{g}_3 =c \left(
\begin{array}[c]{ccc}
-1 & 1 & 0\\ 0& -1& 1\\ 1 &0 & -1 \\
\end{array}
\right) \mbox{ or } \Delta \hat{g}_3 =c \left(
\begin{array}[c]{ccc}
-1 & 0 & 1\\ 1& -1& 0\\ 0 &1 & -1 \\
\end{array}
\right)
$$ where $c>0$ is a constant. We call such contributions, which
involve a closed loop of at least three terminals (here, $1\rightarrow
2 \rightarrow 3$ or $1 \rightarrow 3 \rightarrow 2$ ) in an order
selected by the magnetic field orientation, a ``chiral'' contribution.

This approach can be straightforwardly generalized to $N_T >3$. In
cases with broken time-reversal symmetry, we expect to have some
symmetric and some chiral contributions. To be more precise, we define
the matrix
\begin{equation} 
\label{e4.6}
l(a,b)|_{\alpha\beta} = \delta_{\alpha a} \delta_{\beta b}
-\frac{1}{2} \delta_{\alpha a} \delta_{\beta
a}-\frac{1}{2}\delta_{\alpha b}\delta_{\beta b}
\end{equation} 
which contributes a unit to the off-diagonal element $g_{ab}$. For any
ordered sequence of terminals $a_1, a_2, ..., a_n$, we define the
matrix:
\begin{equation}
\label{e4.5}
\hat{r}(a_1,\cdots,a_n)
=\hat{l}(a_1,a_2)+\hat{l}(a_2,a_3)+\dots+\hat{l}(a_n,a_1).
\end{equation} 
Any such $\hat{r}$ matrix satisfies all the constraints of
Eq.~(\ref{diag}).  With this notation, and assuming that the magnetic
field is such as to select the first form of $\Delta \hat{g}_3$, we
can rewrite:
%%%%%%%%%%%%%%%%%%%%%%%%%%%%%% EQUATION %%%%%%%%%%%%%%%%%%%%%%%%%%%%%%
\begin{equation}
\label{nn10}
\hat{g}_3 = G_{12}\hat{r}(1,2) + G_{13}\hat{r}(1,3) +
G_{23}\hat{r}(2,3)+c \hat{r}(1,2,3)
\end{equation}
%%%%%%%%%%%%%%%%%%%%%%%%%%%%%%%%%%%%%%%%%%%%%%%%%%%%%%%%%%%%%%%%%%%%%%

Any $N_T\times N_T$ conductance matrix $\hat{g}$ can be decomposed in
a similar fashion, into a sum of symmetric (``resistances'') and
chiral contributions:
\begin{equation}
\label{e4.4}
\hat{g} = \sum c_{a_1,\cdots,a_n}\hat{r}(a_1,\cdots,a_n).
\end{equation} 
where $c_{a_1,\cdots,a_n}$ are positive numbers. The symmetric part is
simply $\sum_{a<b}^{} G_{ab}\hat{r}(a,b)$, where
$G_{ab}=\min(g_{ab},g_{ba})$. This leaves at most $N_T(N_T-1)/2$
finite, positive off-diagonal matrix elements, which can be grouped
into a sum of ``chiral'' contributions. To do this, start with the
smallest non-zero matrix element left, say $g_{ab}-G_{ab}$, then draw
a loop $a \rightarrow a_1 \rightarrow \dots \rightarrow a_n
\rightarrow b$ with bonds only connecting pairs of contacts sharing
non-zero matrix elements. We can now separate a contribution
$(g_{ab}-G_{ab}) \hat{r}(a,a_1,...,a_n,b)$, insuring that the number
of zero off-diagonal matrix elements of the remaining matrix has
increased by at least one, while all the finite off-diagonal matrix
elements are still positive. The procedure is repeated until the
decomposition is completed.

\subsection{Six-terminal geometry for the IQHE}

While the decomposition of Eq.~(\ref{e4.4}) is very general, which of
the decomposition terms are important depends, of course, on the
physical system considered. We now return to the problem of interest
to us, namely the $6 \times 6$ matrix $\hat{g}(\nu)$ describing the
contribution to the total conductance of the LL hosting the Fermi
energy. From now, $0<\nu<1$ stands for the filling factor of the LL
hosting the Fermi energy. The total filling factor is $n+\nu$, where
$n$ is the number of filled LLs.

At low-$\nu$, all states of the LL hosting $E_F$ are localized and
transport can only occur through tunneling. Tunneling occurs with
equal probability in both directions. Thus, on very general grounds,
here we expect the conductance matrix to be symmetric, with small
off-diagonal elements since tunneling probabilities are small.

\begin{figure}[t]
\centering \includegraphics[width=0.9\columnwidth]{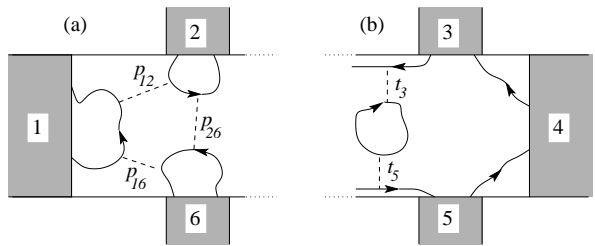}
\caption{ A semi-classical illustration of possible transport
        between terminals, for (a) low filling factors $\nu\rightarrow
        0$. In this case, all
        states in the LL hosting the Fermi level are localized, and
        transport can only occur by tunneling.
        (b) at high filling factors $\nu\rightarrow 1$, the edge states
        are established and dominate the transport. Jain-Kivelson tunneling through
        localized states  inside the sample can create short chiral currents. }
\label{fig16}
\end{figure}

Let us consider this case in more detail. In Fig.~\ref{fig16}(a), we
sketch some possible routes for charge transport between three terminals, at
low-$\nu$. The wave-functions are drawn in a semi-classical manner,
following equipotentials of the disorder potential in the direction
determined by the sign of $B$. This does not mean that our arguments
 only hold in the semi-classical regime. They are general
and hold in the quantum regime -- we just do not know how to draw
quantum mechanical wave-functions.

Most of the electrons injected through any of the terminals will be
scattered back. However, with a small probability $p_{12}$, electrons
injected from terminal 2 can tunnel to near terminal 1. Another
possible route of scattering from 2 into 1, is for electrons to first
tunnel to a state near 6 (probability $p_{26}$) and from there to near
1 (probability $p_{16}$). The total probability for this process is
$(1-p_{12})p_{26}p_{61}(1-p_{12})$. The electron can, however, make
any number of closed loops between finally entering in 1, and so the
total probability to arrive from 2 to 1 is
$$ p_{2\rightarrow 1} = p_{12} +
\frac{(1-p_{12})p_{26}p_{61}(1-p_{12})}{1-p_{12}p_{26}p_{61}}
$$ Similarly, an electron injected through 1 can scatter into 2 either
directly [with probability $(1-p_{16})p_{12}(1-p_{26})$] or can make
any number of loops, yielding
$$ p_{1\rightarrow 2} =
\frac{(1-p_{16})p_{12}(1-p_{26})}{1-p_{12}p_{26}p_{61}}
$$ The other probabilities $p_{1\rightarrow 6}, p_{6\rightarrow 1},
p_{2\rightarrow 6}$ and $p_{6\rightarrow 2}$ can be calculated
similarly.
 
Using $G_{ab} = \frac{e^2}{h}\min( p_{a\rightarrow b}, p_{b\rightarrow
a}) >0$, we find $G_{\alpha\beta} =\frac{e^2}{h}
p_{\alpha\beta}+O(p^3)$, {\em i.e.}  indeed, the largest contribution
to transport between these 3 terminals comes from direct tunneling and
is symmetric. However, the conductance matrix is not fully symmetric,
since $p_{2\rightarrow 1} - p_{1\rightarrow 2} >0$. In fact, one finds
that $p_{2\rightarrow 1} - p_{1\rightarrow 2}=p_{6\rightarrow 2} -
p_{2\rightarrow 6}= p_{1\rightarrow 6} - p_{6\rightarrow 1} = c =
p_{12}p_{26} +p_{12}p_{16} +p_{26}p_{61} +O(p^3) $, showing also the
appearance of a very small chiral contribution. Thus, the contribution
of these processes to the conductance matrix is precisely of the
expected form of Eq.~(\ref{nn10}).  This derivation completely ignores
interference effects between different scattering paths, and therefore
is valid only in the presence of significant dephasing. Our numerical
simulations, on the other hand, assume full coherence between all
electron wave-functions (there is no inelastic scattering in our
Hamiltonian). In such a case, one should sum the amplitudes of
probabilities for various processes, and then square its
modulus to find the total probability. The derivation for this case is very similar
to the above one (also see Ref.~\onlinecite{Zhou04c}). The main
contribution to the off-diagonal conductance matrix elements are still
due to the direct tunneling, {\em e.g.} $G_{12} = \frac{e^2}{h}
|t_{12}|^2$, where $t_{12}$ is the amplitude of probability to tunnel
from 1 to 2 (so that $|t_{12}|^2=p_{12}$). The  difference is that
the chiral current is now of order $|t|^3$, not $p^2=|t|^4$ as when the
interference is ignored. The reality is in between, since in the real
samples there is some amount of decoherence. Irrespective of how much
decoherence there is, a predominantly symmetric conductance matrix is
inevitable if tunneling is the only means of charge transport.

 Of course, in order to derive an expression for the entire $6\times6$
conductance matrix, we have to also consider tunneling to the other 3
terminals in all possible combinations. It should be apparent that as
long as all $p \ll 1$, the only effect of that is to add symmetric
terms of order $p$ between all pairs of terminals connected by direct
tunneling, and much smaller chiral terms, of order $p^2$ or $|t|^3$,
for various closed loops.  This explains why for low-$\nu$, $\hat{g}$
is symmetric with small off-diagonal components, as indeed confirmed
by the numerical simulations (see Fig.~\ref{fig15}).

At high-$\nu$, on the other hand, the transport mechanism is very
different. As expected on general grounds (and confirmed by the
numerics) edge states are established within the LL once $\nu
>0.5$. With very high probability, electrons are transported through
the edge states to the next terminal $ \alpha+1 \rightarrow \alpha$.
In the limit $\nu \rightarrow 1$ we expect and find $\hat{g}
\rightarrow \frac{e^2}{h}\hat{r}(1,2,3,4,5,6) = \hat{g}^{(0)}$.

For intermediate $\nu$, however, shorter chiral loops containing edge
states can be established through tunneling, as sketched in
Fig.~\ref{fig16}(b). Assume that an electron leaving contact 3 
can tunnel with amplitudes of probability $t_3$ and $t_5$ to and out
of a localized state inside the sample, to join the opposite edge
state and enter 5. This is precisely the Jain-Kivelson
phenomenology.\cite{Kivelson88} Summing over all possible processes
and ignoring decoherence, we find their result\cite{Kivelson88}
\begin{equation}
\label{e4.9}
p_{3\rightarrow 5} =\frac{h}{e^2} g_{53} =
\left|\frac{t_3t_5}{1-r_3r_5\exp(i\phi/\phi_0)}\right|^2
\end{equation}
where $\phi$ is the flux enclosed by the localized state, $\phi_0=h/e$
is the unit of flux, and $|r_3|= \sqrt{1-|t_3|^2}, |r_5|=
\sqrt{1-|t_5|^2}$ are the amplitudes to avoid the corresponding
tunneling. On the other hand, $p_{5\rightarrow 3} =0$, since in this
scenario, no electron leaving 5 enters 3. Thus, in this case there is
no symmetric term proportional to $\hat{r}(3,5)$, only a chiral loop
term $g_{53}\hat{r}(3,4,5)$. Physically, this term represents the
backscattered current of the Jain-Kivelson model.\cite{Kivelson88}
Other shorter chiral loops can be established by tunneling between
other pairs of edge states, so in the high-$\nu$ limit we expect the
conductance matrix to be a sum of such chiral loops. Again, inclusion
of partial or total dephasing does not change this conclusion.

This analysis has reconfirmed our assertion that the conductance
matrix can be decomposed into a sum of symmetric terms and chiral
terms. For this particular system, we have now shown that at
low-$\nu$, the symmetric terms are the dominant (although small)
contribution, while at high-$\nu$, chiral loops are the dominant
contribution. Near half-filling, we expect both types of terms to be
important.  Consider then the general form:
\begin{widetext}
\begin{eqnarray}
\label{e4.10}
\nonumber \hat{g}&=& n \hat{g}^{(0)}+
 G_{12}\hat{r}(1,2)+G_{16}\hat{r}(1,6)+G_{26}\hat{r}(2,6)+
 G_{34}\hat{r}(3,4)+G_{35}\hat{r}(3,5)+G_{45}\hat{r}(4,5)+ \\ &&
 +\frac{e^2}{h}\left[c_0\hat{g}^{(0)}+c_1\hat{r}(1,2,6)+
 c_2\hat{r}(2,3,5,6)+c_3\hat{r}(3,4,5) +c_4\hat{r}(1,2,3,5,6)+
 c_5\hat{r}(2,3,4,5,6)\right].
\end{eqnarray} 
\end{widetext}
The first term is the contribution of the $n$ completely filled lower
LLs, while all other terms describe transport in the LL hosting
$E_F$. Eq.~(\ref{e4.10}) is not the most general possible
decomposition -- that would have 15 symmetric and 10 chiral terms. In
Eq.~(\ref{e4.10}) we assume no tunneling (no symmetric terms) between
the left and right sides of the sample. The largest such neglected
terms are $G_{23} \hat{r}(2,3)$ and $G_{56}
\hat{r}(5,6)$. Numerically, we find $G_{23}, G_{56}
<10^{-4}\frac{e^2}{h}$ [see Fig.~\ref{fig15}, where
$G_{23}=\min(g_{23},g_{32})$, etc.]. We analyze the influence of these
neglected terms on the resistance fluctuations in the next
section. The 4 chiral loops neglected all have one ``diagonal'' link
between 2 and 5, or between 3 and 6. As explained, chiral loops are
important at larger filling factors, because of Jain-Kivelson 
scattering between opposite edge states. Such scattering does not
mediate direct transport between $2\leftrightarrow 5$ or
$3\leftrightarrow 6$ (see discussion below), so these terms can be
safely neglected.

\begin{figure}[bh]
\centering \includegraphics[width=\columnwidth]{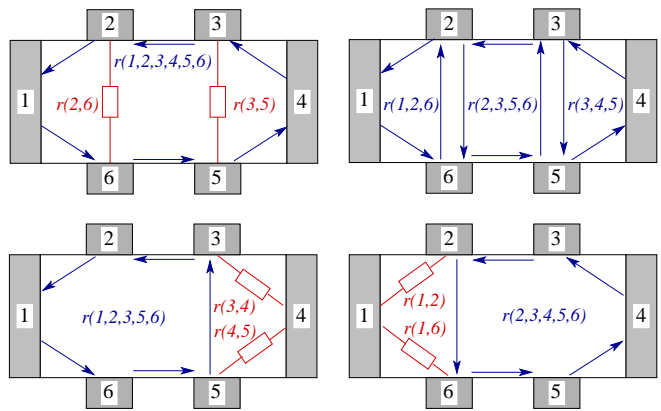}
\caption{Terms contained in the general decomposition of \Eref{e4.10}. Symmetric
  terms are represented as resistors, while chiral terms are
  represented as closed directed loops. }
\label{fig17}
\end{figure}

The model of \Eref{e4.10} is thus a parameterization of the 6$\times 6$
matrix $\hat{g}$ with 12 independent parameters. At low $\nu$, the
chiral terms $c_{i} \rightarrow 0$ and the decomposition is dominated
by the symmetric terms. At high $\nu$, the symmetric terms vanish
$G_{\alpha\beta} \rightarrow 0$ while the chiral terms are important,
in particular $c_0 \rightarrow 1$. Near half-filling, all terms are
likely to contribute. \Eref{e4.10} thus covers all possible filling
factors $0<\nu<1$.  Figure~\ref{fig17} sketches the elements contained in
this model. The closed directed loops indicate chiral terms while the
resistors indicate the existence of symmetric terms. Resistors
suggest dissipation; as we show shortly, they play a role only during
transitions between QHE plateaus. Close to integer filling, where the
Hall conductances are quantized and the sample is disipationless,
$\hat{g} \rightarrow n \hat{g}^{(0)}$ and all symmetric
(``resistors'') terms vanish, as do the shorter chiral loops.

\subsection{The correlations of the resistance fluctuations}
 
With Eq.~(\ref{e4.10}), the equations $\hat{I}_{14}=\hat{g}\cdot
\hat{V}$ and $\hat{I}_{63}=\hat{g}\cdot \hat{V'}$ can be solved
analytically and the various Hall, longitudinal and two-terminal
resistances can be calculated in terms of these 12 parameters.  The
complete solutions are very long and we do not list them here. 

The following identity is found to hold:
\begin{equation}
\label{e4.11}
 R_{14,63}=R_{63,63} = \frac{h}{e^2} \frac{1}{n + c_0 + c_2 + c_4
 +c_5}.
\end{equation}
Since $R_{63,63}=R_{2t}$, whereas $R_{14,63} = R^H_{14,62}+R^L_{14,23}
=R^L_{14,65}+ R^H_{14,53}$, this means that $R_{2t} = R^H + R^L $
irrespective of the value of the 12 parameters. In other words, this
identity is obeyed for all $\nu$, which explains why it is observed in
both experiment and simulation.  (This identity is not observed in
experiment for low magnetic fields.  This can be ascribed to 
deviations from the IQHE regime, where some of the approximations we
make here -- ignoring the LL mixing, for example --
fail.)

In \Eref{e4.11}, $n + c_0 + c_2 + c_4 + c_5$ is the total chiral
current flowing along the $6\rightarrow 5$ and $3\rightarrow 2$ edges.
As discussed, at low-$\nu$ the chiral currents in the LL hosting $E_F$
are vanishingly small: $c_i=0$, $i=1,...,6$ (there are no edge states
established yet, and tunneling contributions to chiral currents are of
order $t^3$ or less, as showed in the previous section. Below $\nu_c$, all $t < 10^{-2}$,
see Fig.~\ref{fig15}). It follows that at low-$\nu$, $R^L + R^H =
h/(ne^2)$, explaining the perfect correlations in the fluctuation
patterns of the two resistances, seen experimentally and numerically.

\begin{figure}[t]
\centering \includegraphics[width=0.6\columnwidth]{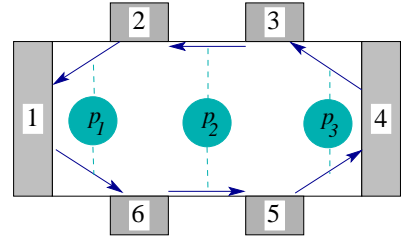}
\caption{Possible Jain-Kivelson processes in the sample for high-$\nu$
region. }
\label{fig4.34}
\end{figure}

In the high-$\nu$ regime the transport is dominated by the chiral
currents created by tunneling between opposite edge states, through
localized states inside the sample.\cite{Kivelson88} The most general
possible situation is sketched in Fig.~\ref{fig4.34}.  $p_1$, $p_2$,
and $p_3$ are the total probabilities for tunneling between the pairs
of edges, summing over all the possible tunneling processes through
all available localized states in the sample. Each such contribution
fluctuates strongly as $B$ (or $\nu$) is changed, since the magnetic
flux enclosed by various localized states changes significantly [see
Eq.~(\ref{e4.9}) for the simplest possible expression for such a
probability]. The various conductance matrix elements can be
simply read off this figure; for example, $\frac{h}{e^2}g_{12} =
p_{2\rightarrow 1} = 1-p_1$. After adding the contribution of the $n$
filled LL, the total conductance matrix can be written as a sum of
chiral terms, consistent with \Eref{e4.10}:
\begin{eqnarray}
\label{e4.12}
\nonumber &&\hat{g}=
 (1-p_1-p_2-p_3)\hat{g}^{(0)}+\frac{e^2}{h}p_2\left[\hat{r}(1,2,6) +
 \hat{r}(3,4,5)\right] \\ && +\frac{e^2}{h}\left[p_3
 \hat{r}(1,2,3,5,6) + p_1 \hat{r}(2,3,4,5,6)\right] +n\hat{g}^{(0)} .
\end{eqnarray} 
In this case, the equation $\hat{I}_{14}=\hat{g}\cdot \hat{V}$ is
trivial to solve. We find:
\[ R^H_{14,62} = R^H_{14,53} = \frac{h}{(n+1)e^2},\] 
i.e. the Hall resistances are precisely quantized, whereas
\[ R^L_{14,23} = R^L_{14,65} = \frac{h}{(n+1)e^2}\cdot \frac{p_2}{n+1-p_2}.\] 
Note that the result only depends on $p_2$, {\em i.e.} on tunneling
between edge states in the central part of the device, between the 4
voltage terminals. This is further justification that a short sample
is sufficient for the numerical simulation. Since $p_2$ has a strong
resonant dependence on $E_F$ (or $\nu$), it follows that the two $R^L$
fluctuate strongly, but with identical patterns. This is precisely
what is observed in experiment and numerics, on the high-$\nu$ side of
the transition (see Figs.~\ref{fig14} and \ref{fig12}) and supports
our assertion that the fluctuations in this regime are caused by
Jain-Kivelson tunneling.  In particular, if $n=0$ (transition inside
the spin-up LLL), $R^L$ can be arbitrarily large when $p_2\rightarrow
1$ (see Fig.~\ref{fig14}). In higher LLs, the amplitude of
fluctuations of $R^L$ is $h/[n(n+1)]e^2$ or less, as observed both
experimentally and in our simulations (see Fig.~\ref{fig12}).

These results are very interesting because they show that one does not
actually need to {\em know} the conductance matrix in order to understand
what correlations might exist between various resistances. All that is
needed is to have some idea of its general structure, which can be
inferred based on physical arguments. At low $\nu$ the conductance
matrix is the sum between $n\hat{g}^{(0)}$ and a symmetric matrix
describing tunneling between various terminals (ignoring tunneling
between the left and right sides, for the time being). The solutions
of $\hat{I}_{14} = \hat{g}\cdot \hat{V}$ for such a matrix always
satisfy $R^L+R^H=h/(ne^2)$, no matter what are the off-diagonal values
of the symmetric component. It follows that this identity must be
obeyed in experiments as long as the conductance matrix has this form,
{\em i.e.} for filling factors low-enough that tunneling is the
dominant transport mechanism in the LL hosting $E_F$. On the other
hand, for high filling factors, physical arguments based on the
appearance of the edge states and the Jain-Kivelson phenomenology
lead to the conclusion that the general form of the conductance matrix
is like in Eq.~(\ref{e4.12}). In this case, we find that irrespective
of the values of the parameters $p_1, p_2$ and $p_3$, the $R^H$ are
quantized while the two $R^L$ fluctuate with identical
patterns. Finally, to cover all possible $\nu$, we must include in
$\hat{g}$ both types of allowed symmetric and chiral terms. In this
case, we find that the identity $R_{2t}=R^H+R^L$ holds, irrespective
of the values of the various parameters. Such arguments can be
straightforwardly generalized to geometries with any number of
terminals, allowing one to easily test in what cases are correlations
expected on such general grounds. The behavior when the sign of $B$ is
changed can also be understood easily, as we show now.

\subsection{Changing the orientation of the magnetic field}

If $B$ changes sign, the Onsager reciprocal relation\citep{Baranger89}
reads
\[ \hat{g}(-B) = [\hat{g}(B)]^T . \]   
That this must be so is obvious for our generic conductance matrix of
Eq.~(\ref{e4.10}): the time-reversal symmetric tunneling is not
affected by this sign change, but the flow of the chiral currents is
reversed (equivalent with taking the transpose).  The solutions of
$\hat{I}_{14} = \hat{g}(-B) \cdot \hat{v}$ are then related to the
solutions of $\hat{I}_{14} = \hat{g}(B) \cdot \hat{V}$ by $v_2 = V_6$,
$v_3 = V_5$, $v_5 = V_3$ and $v_6 = V_2$, provided that the same index
exchanges $ 2 \leftrightarrow 6$, $ 3 \leftrightarrow 5$, are
performed for all $G_{ab}$.  Terms not invariant under this
transformation are $G_{12}$, $G_{16}$, $G_{43}$, and $G_{45}$. In the
experimental setup, these four terms must be very small, due to the
long distance between source and drain, and their nearby contacts. (In
the simulation, these terms are sometimes not negligible because the
simulated sample is rather short). If we set these 4 terms to zero and
keep only the largest symmetric terms $G_{26}$ and $G_{35}$ in
Eq.~(\ref{e4.10}), we find that $R^L_{14,23}(B)= R^L_{14,65}(-B)$ and
vice versa. In other words, the fluctuation pattern of one $R^L$
mirrors that of the other $R^L$ when $B \rightarrow -B$, as observed
experimentally.\cite{Peled04} Small violations of this symmetry
observed experimentally at low-$\nu$, are likely due to the
perturbative corrections from the very small, non-invariant tunneling
contributions proportional to $G_{12}-G_{16}$ and
$G_{43}-G_{45}$. This is confirmed in the next section.

\subsection{Small corrections to  correlations and symmetry}

The most significant terms neglected in the general decomposition of
\Eref{e4.10} are $G_{23} \hat{r}(2,3)$ and $G_{56}\hat{r}(5,6)$, which
are due to tunneling between the neighboring voltage probes.  All
other tunneling terms neglected are smaller than these, because they
are between contacts further apart. Remember that even these two terms
are very small; numerically we found that they are of order
$10^{-4}\frac{e^2}{h}$ or less (see Fig. \ref{fig15}). We now
investigate whether the inclusion of these terms violates
significantly the various correlations established in their
absence. For simplicity, in the rest of this section we set $e^2/h
=1$, in other words we measure all conductances in $e^2/h$  units.

We first investigate the effects of adding the $G_{23}$ and $G_{56}$
terms on the low-$\nu$ correlations $R^H+R^L=1/n$. We set all $c_i=0$,
since at low-$\nu$ the chiral terms are negligible compared to the
symmetric terms. To first order perturbation in $G_{23}$ and $G_{56}$,
we find:
\begin{eqnarray}
\nonumber R_{14,63} - \frac{1}{n} = { [n(G_{26}+G_{12}) + F_1]G_{23}
\over n^2[n F_2 +F_1]}&&\\\label{e4.13} + {[n(G_{35} + G_{45})+ F_3]
G_{56} \over n^2[nF_4+F_3]} + \cdots,&&
\end{eqnarray}
where
\[ F_1 = G_{16}G_{26}  +G_{12}G_{16} +G_{12}G_{26},\; 
F_2 = G_{16}+G_{26}+ G_{12},\]
\[ F_3 = G_{34}G_{45} +G_{34}G_{35}+G_{35}G_{45},\; 
F_4 = G_{34}+ G_{35}+G_{45}. \] In the given experimental geometry, we
expect that $G_{12}, G_{16} \ll G_{26}$ and $G_{34}, G_{45} \ll
G_{35}$, since terminals 1 and 4 are very far from 2 and 3,
respectively 5 and 6, and thus the tunneling probabilities must be
much smaller. Within this limit, the previous expression can be
further simplified to:
$$ R_{14,63} - \frac{1}{n} \approx \frac{G_{23}+G_{56}}{n^3}
$$ In other words, the identity $R^H+R^L = R_{14,63} = 1/n$ should
indeed be obeyed with high accuracy, as long as the direct tunneling
between 2 and 3, respectively 5 and 6, as well as the chiral
contributions, are indeed small.

We now consider the effect of adding the $G_{23}$ and $G_{56}$ terms
on the identity $R^L+R^H= R_{2t}$. After expanding to first order in
these two quantities, we find:
\begin{equation}
\label{e4.14}
R_{14,63}-R_{63,63} = - { G_{45}G_{56} \over A_1} - {G_{12}
  G_{23}\over A_2} +\cdots .
\end{equation}
where we define $C_T = \sum_{i=0}^{5}c_i$ and
$$ A_1 = (n+C_T-c_1+G_{35})(n+C_T-c_1-c_3)(n+c_0+c_3+c_5) $$
$$ A_2 = (n+C_T-c_3+G_{26})(n+C_T-c_1-c_3)(n+c_0+c_1+c_4) $$ Since
$G_{45}$, $G_{56}$, $G_{12}$, and $G_{23}$ are all small numbers, we
have $A_1, A_2 \sim n^3 \geq 1$, and therefore this correction is also
small for all values of $\nu$.

Finally, we investigate how the $B$-reversal symmetry is perturbed by
a small imbalance between $G_{34}$ and $G_{45}$, respectively $G_{12}$
and $G_{16}$.  As stated before, $R^H(-B)$ and $R^L(-B)$ are
calculated with $\hat{g}^T$.  For simplicity, here we take
$G_{23}=G_{56}=0$, and find:
\begin{subequations}
\label{e4.15}
\begin{eqnarray}
\label{e4.15a}
R^L_{14,65}(B)-R^L_{14,23}(-B) = {G_{34}-G_{45} \over B_1},&&\\
\label{e4.15b}
R^L_{14,23}(B)-R^L_{14,65}(-B) = {G_{12}-G_{16} \over B_2}.&&
\end{eqnarray} 
\end{subequations}
where
$$ B_1=G_{34}G_{45} + (n+c_0+c_3+c_5+G_{34}+ G_{45})(n+C_T-c_1+G_{35})
$$
$$ B_2=G_{12}G_{16} + (n+c_0+c_1+c_4+G_{16}+G_{12}) (n+C_T-c_3+G_{26})
$$

The largest contributions to $B_1$ and $B_2$ come from $n, G_{35}$ and
$G_{26}$ and explain why these corrections to perfect symmetry of
$R^L_{14;65}(B)$ and $R^L_{14;23}(-B)$ are small. In fact, we can see
that the larger the various chiral terms, the smaller these
corrections should be. This agrees with the experimental data, where
the largest violations of this symmetry are observed between the
neighboring large fluctuation peaks in $R_{14;63}$, on the low-$\nu$
side of the central regime. From \Eref{e4.11}, we know that those
sharp peaks are caused by fluctuating chiral components of the
conductance matrix, such as $c_2, c_4,$ and $c_5$. From
Eqs. (\ref{e4.15}), we see that when these parameters are large (peak
value in $R_{14;63}$), the difference between $R^L_{14;65}(B)$ and
$R^L_{14;23}(-B)$ is suppressed. When $R_{14;63}$ is closer to $1/n$,
however, the chiral terms are smaller and the corrections to the
difference becomes more noticeable.

In all these correction terms, powers of $n$ appear in the
denominator. Since $n$ is the number of underlying filled LLs, these
corrections should be smaller in higher LLs and therefore the
symmetries and correlations should be easier to observe. They have
been experimentally observed in higher LLs,\cite{Peled04} but they are
not so clear as in the first plateau-to-plateau transition. There are
two reasons for this. First, the sample quality is rather poor and
only the first transition is clearly observed.\cite{Peled03b,Peled04}
Secondly, a smaller magnetic field is needed to reach higher LLs.  As
the cyclotron energy is reduced, inter-Landau level mixing and other
effects which we did not consider in these simulations may start to
become important.
 
\section{Summary and discussion}
\label{ssum}

To summarize, in this study we show that first-principles simulations
of the IQHE in mesoscopic samples, based on the multi-terminal
Landauer formalism appropriate for non-interacting electrons,
recapture all the non-trivial correlations and symmetries recently
observed experimentally. Moreover, we explain how these correlations
and symmetries are direct consequences of the general allowed
structure of the conductance matrix.

Similar to the experiments, we find that the IQHE transition in higher
LLs is naturally divided into three regimes. On the low-$\nu$ side of
the transition, the LL hosting $E_F$ is insulating. If tunneling
between left and right sides is also small, the fluctuations of pairs
of resistances are correlated with excellent accuracy, i.e. $R^H +R^L
= h/ne^2$. This condition is obeyed if the typical size of the
wave-function (localization length) is less than the distance between
contacts 2 and 3.  When the localization length becomes comparable to
this distance, edge states begin to be established and the correlation
between $R^L$ and $R^H$ is lost.  On the high-$\nu$ side of the
transition, the edge states are established and responsible for most
of the charge transport. However, localized states inside the sample
can help electrons tunnel between opposite edges, leading to
back-scattering like the Jain-Kivelson model. In this case, our
simulations show that the two $R^L$ fluctuate with identical patterns,
while the $R^H$ are quantized. Tunneling between opposite edges is
likely only if the typical size of the wave-functions is slightly
shorter than the distance between opposite edges.  It is then apparent
that the central regime in Figs. \ref{fig11} and \ref{fig12}
corresponds to the so-called ``critical region'', where the typical
size of the electron wave-function is larger than the sample size
(distance between contacts 2 and 3, at low-$\nu$, or between 2 and 6
at high-$\nu$).  In these mesoscopic samples, the voltage probes act
as markers on a ruler, measuring the size of the wave-functions at the
Fermi energy.  To our knowledge, this is the first time when the
boundaries of the critical region are pinpointed experimentally.  This
opens up exciting possibilities for experimentally testing the
predictions of the localization theory of IQHE.

Conductance fluctuations at the IQHE transitions have been studied
before by several authors. Cho and Fisher~\cite{Fisher97} and Wang
{\em et al.}\cite{Wang96} have focused on the two-terminal
conductance, with numerical simulations based on Chalker and
Coddington's network model.\cite{Chalker88} Ando has numerically
computed conductances for two- and four-terminal samples,\cite{Ando94}
using the B\"uttiker-Landauer formalism, but in the Green's function
formulation that requires the discretization of the sample to a
lattice model.\cite{Ando91,Datta97} Their results are consistent with
ours, showing fluctuations in the resistances which are random,
sample-dependent, and of order of $h/e^2$. St\v{r}eda, Kucera and
MacDonald\cite{MacDonald87} were the first to predict the relations
$R^H+R^L = h/(ne^2)$ and $R_{2t} = R^H+R^L$, using an analytical
analysis of the Landauer formula for a ``two-terminal'' sample, but which
has the upper and  lower sides of each terminal held at different voltages. In
effect, their current source and drain (leads 1 and 4) are 
moved to infinity, and connected to the central sample through edge
states only. This is a special case of our general model of
Eq.~(\ref{e4.10}). B\"uttiker~\cite{Buttiker86} presented a detailed
analysis of four-terminal systems, similar to our analysis of the
six-terminal conductance matrix. He  predicted the symmetry relations
between conductance measurements which exchange the role of the
voltage and current leads.

\begin{figure}[b]
\centering \includegraphics[width=0.5\columnwidth]{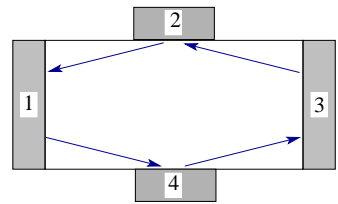}
\caption{Edge states in a four-terminal measurement. Leads 1 and 3
  are the current source and drain, while leads 2 and 4 are the
  voltage leads. }
\label{fig28}
\end{figure}

However, it is essential to emphasize that neither two, nor
four-terminal samples (with terminals at well defined voltages)
can simultaneously measure both a Hall and a longitudinal
resistance. This is obvious for a two-terminal measurement, which can
 measure a single two-terminal conductance $R_{12,12}$. For a
four-terminal geometry (see Fig. \ref{fig28}), consider the case of
$n$  filled Landau levels, $\nu=n$. We know that each LL will
transport charge through a set of edge states, such as those sketched
in Fig. \ref{fig28}. The conductance matrix in this case is:
$$
\hat{g}_4 = n\frac{e^2}{h}
\left(
\begin{array}[c]{cccc}
-1 & 1& 0& 0\\
0 & -1& 1& 0\\
0 & 0& -1& 1\\
1 & 0& 0& -1\\
\end{array}
\right)
$$
Taking leads 1 and 3 as the current source and drain,
$\hat{I}=(-I,0,I,0)$ and assuming that lead 3 is grounded, one can
solve $\hat{I}=\hat{g}_4\cdot \hat{V}$ for the other 3
voltages. One  finds trivially that $V_1=V_4=I h/(ne^2)$ and $V_2=V_3=0$. Thus,
even though at first sight one might assume that one can measure a Hall resistance
$R^H=(V_4-V_2)/I$  and
a longitudinal resistance  $R^L=(V_1-V_3)/I$, it turns out that in fact
both measurements give  resistances exhibiting plateaus at quantized values
$h/(ne^2)$. In other words, both results are related to a Hall resistance (a
longitudinal resistance should be zero when $\nu=n$. Note that at
transitions between the plateaus, the two resistances need not be equal, since
the partially filled LL will introduce other off-diagonal matrix
elements. Using the type of analysis we introduced in Section
\ref{sglparam}, one could now easily study what types of correlations
might be possible in such a geometry). If one now
compares this to a 6-terminal case, it becomes obvious that
the four-terminal
resistances have no reason to be related to the  $R^H$ measured in the
six-terminals by experimentalists. Thus, in order to simulate and 
understand the experimental results, it is essential  to use the
six-terminal geometry.  

These considerations show that for mesoscopic samples, a full
specification of the experimental setup is absolutely necessary for
any interpretation of the measured quantities.  For example, the
various correlations and symmetries that we studied in this paper only
appear in the six-lead setup. Other geometries can be analyzed
similarly. A significant result of our work is the proof that an
understanding of various possible (robust) correlations can be obtained based
on  simple arguments regarding the general allowed structure of
the conductance matrix, without need for detailed numerical
simulations. 

Furthermore, we have shown that the full DC response function
(the conductance matrix) of a mesoscopic sample is characterized by
a large number of parameters (12, for our six-terminal geometry, when
we ignore the small tunneling between the left and right-side of the
Hall bar). This is to be contrasted with  macroscopic
samples which only  require two parameters, the Hall and longitudinal
conductivities $\sigma_{xy}$ and $\sigma_{xx}$, to fully characterize
their DC response. A mesoscopic sample 
obviously has more degrees of freedom to display fluctuations of its
 resistances.

\section*{ACKNOWLEDGMENTS}
This research was supported by NSERC and NSF DMR-0213706.  We thank
E. Peled, Y. Chen, R. Fisch, R. N. Bhatt and D. Shahar for helpful
discussions. We also thank Matt Choptuik for providing access to the
von Neumann cluster.

\appendix*

\section{Choosing the right confining potential}
\label{sconf}

The confining potential is needed to define the Hall bar out of the
larger area spanned by each LL Hilbert subspace. It is convenient to
make a symmetric choice. We take the confining potential to be
negative in the region $[-L_x/2,L_x/2)\times(-L_y/2,0)$ (which is thus
the Hall bar) and positive in the remaining region
$[-L_x/2,L_x/2)\times(0,L_y/2)$ (the inverted Hall bar) and such that
$V_c(x,y) = - V_c(x,y+L_y/2)$. On the $y$-edges of the Hall bar we
then have $V_c(x,-L_y/2)=V_c(x,0)=0$. Inside most of the Hall bar, the
confining potential is equal to $-V_{\rm gap}$, where $V_{\rm
gap}>0$ is large enough to confine the electrons to the Hall bar when
disorder is added, for all $E_F <0$.

\begin{figure}[b]
%\centering
\includegraphics[width=0.42\columnwidth]{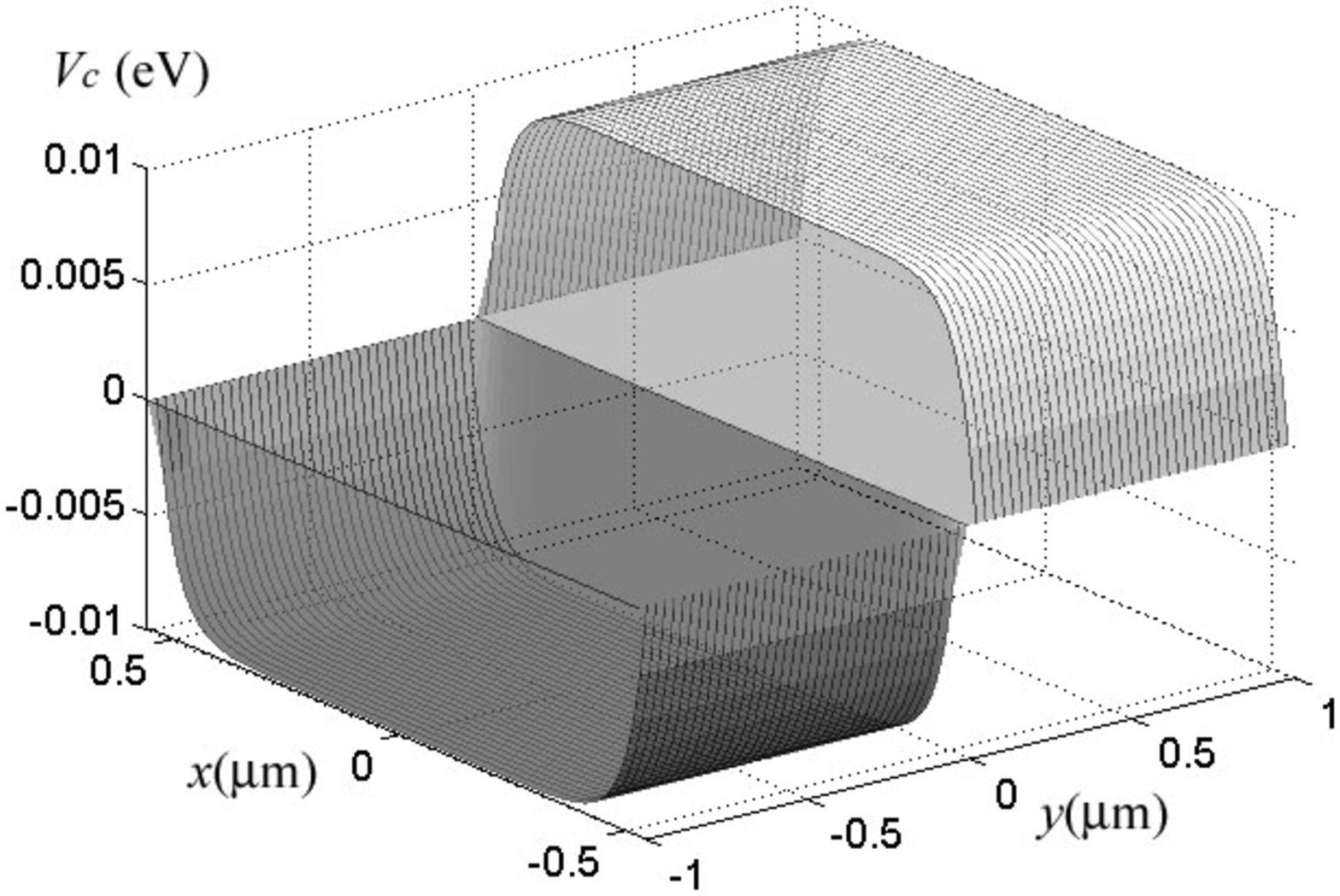}
\includegraphics[width=0.56\columnwidth]{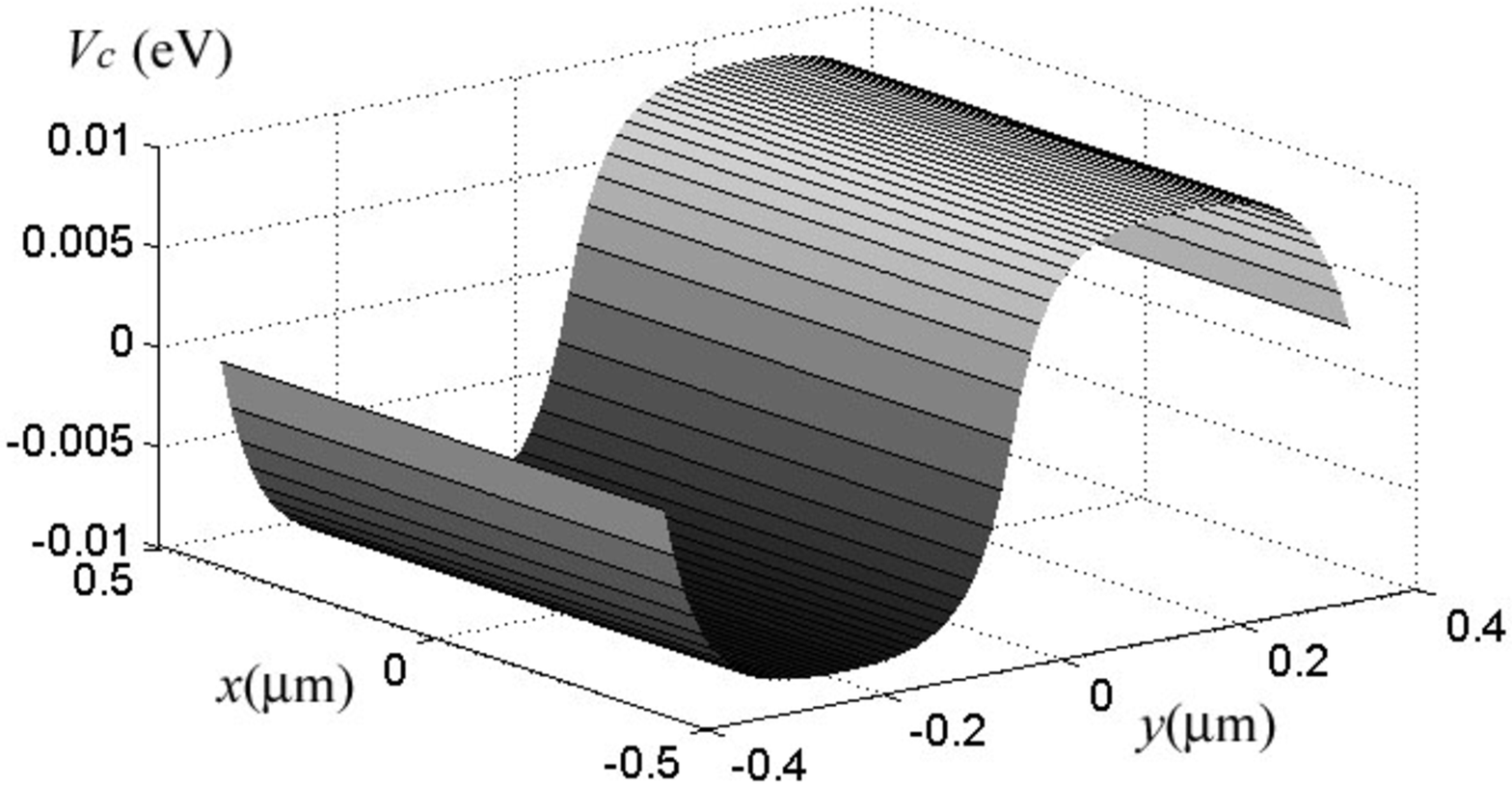}
\caption{Two of the tested confining potentials. Left: This potential
has discontinuous changes at $y=0$ and $y=\pm L_y/2$, and  is smoothly
connected to zero on the $x$-edges. Right: An improved 
potential with open ends and smooth $y$-edges.}
\label{fig4}
\end{figure}

The question is how should $V_c(x,y)$ behave near the Hall bar edges,
at $x=\pm L_x/2$ or $y=-L_y/2,0$. For instance, $V_c$ could vary
smoothly or abruptly from $-V_{gap}$ to 0, as the $y$-edges are
approached. On the $x$-edges, we could take $V_c =0$ or
$-V_{gap}$. Two of the possible choices are shown in Fig.~\ref{fig4}.
The shape of $V_c$ near the edges is critical for the quality of the
simulation, as we show here.

Since little is known about the real shape of the confining potential,
we test the various choices by analyzing the IQHE transition from
$R^H=h/e^2$ to $R^H=h/(2e^2)$, {\em in the absence of disorder}. The
computation is done as described in the text, but with $V_d=0$. We add
a $\hat{g}^{(0)}$ to the calculated $\hat{g}(\nu)$, to account for the
contribution of an underlying, filled LL.  These simulations were
carried on smaller samples, to save CPU time.  The results should
show: (1) a sharp transition of $R^H$ from ${h/ e^2}$ to $h/(2e^2)$;
(2) the built-in particle-hole symmetry of the Hamiltonian, {\em i.e.}
symmetry of the results about $E_F=0$ (this is a consequence of
choosing a symmetric confining potential and symmetric contact states
for the terminals); (3) the two Hall resistances are always equal, and
so are the two longitudinal resistances  (in the absence of disorder,
the potential has rectangular symmetry.)

\begin{figure}[t]
\centering \includegraphics[width=0.9\columnwidth]{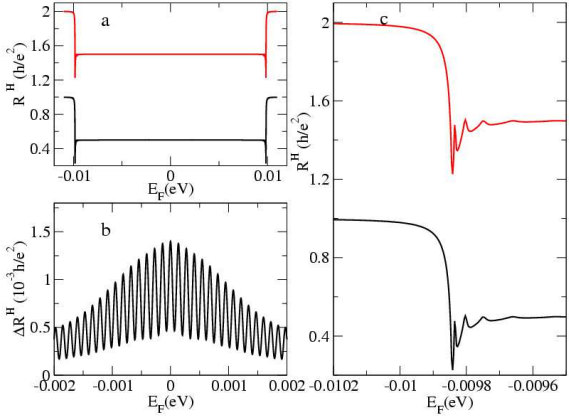}
\caption{Hall resistances calculated for the confining potential shown
in Fig.~\ref{fig4}~(right), $R^H_{14,62}$ (grey) and $R^H_{14,53}$
(black). (a) The overview of the entire energy interval scanned by
Fermi energy $E_F$. Here, $R^H_{14,53}$ has been shifted up by one
unit. (b) Amplified view of the very small oscillations close to
$E_F=0$,  $\Delta R^H = R^H - {h \over 2e^2}$. (c) Amplified view
of the transition from $R^H = {h \over e^2}$ to $R^H = { h \over
2e^2}$. }
\label{fig9}
\end{figure}

\begin{figure}[t]
\centering \includegraphics[width=0.9\columnwidth]{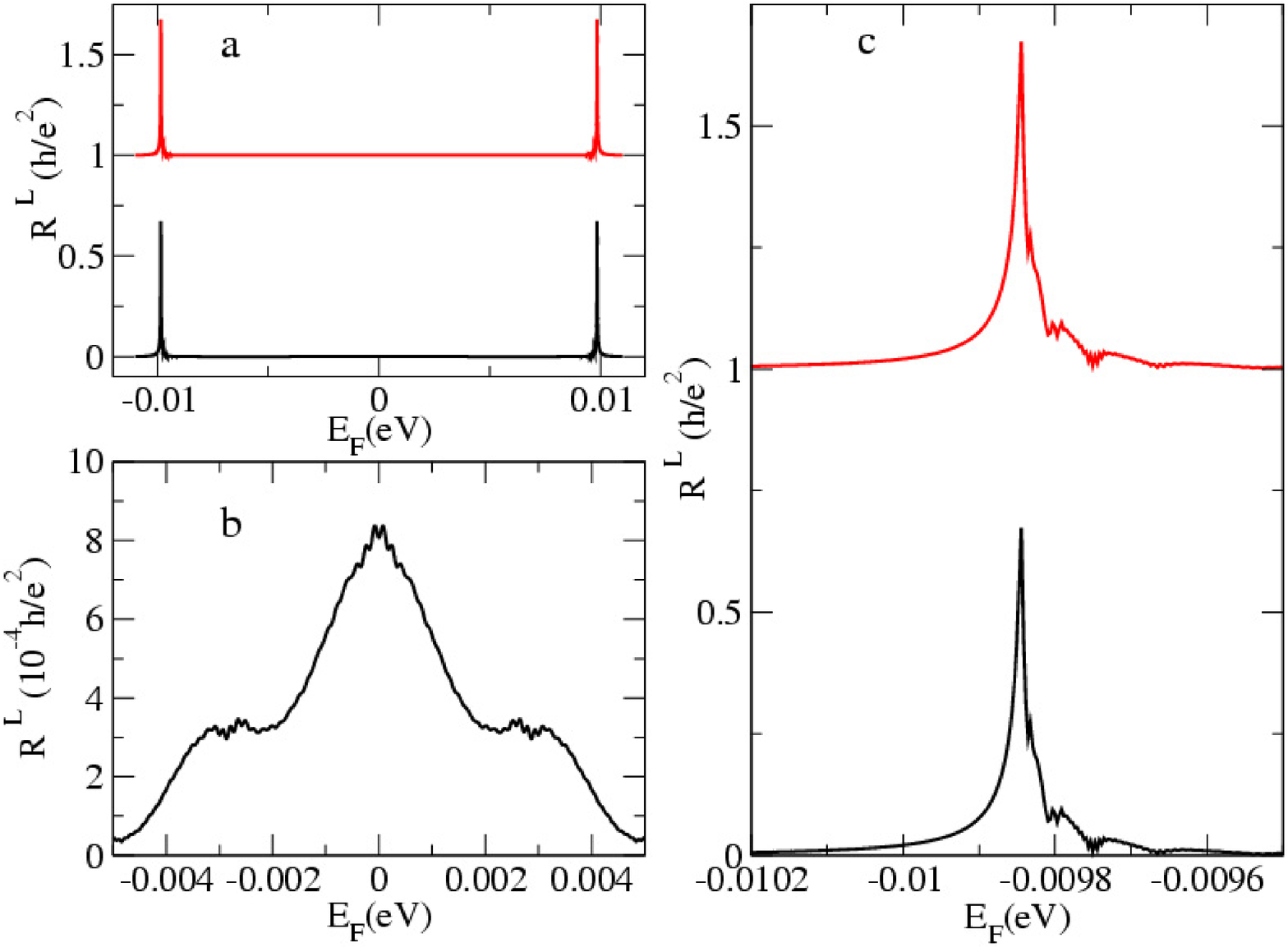}
\caption{Longitudinal resistances calculated for the confining potential shown in
Fig.~\ref{fig4}(right), $R^H_{14,62}$ (grey) and $R^H_{14,53}$
(black). (a) The overview of the entire energy interval scanned by
Fermi energy $E_F$. Here,  $R^H_{14,53}$ has been shifted up by one unit. (b)
Amplified view of the very small variations close to $E_F=0$. (c) Amplified view of the
IQHE transition. }
\label{fig10}
\end{figure}

Figure~\ref{fig4} shows two confining potentials which were used for
testing purposes. Typical results of tests using the potential in the
right panel of Fig.~\ref{fig4} are shown in Fig.~\ref{fig9} and
\ref{fig10}.  We found several features of the confining potential
that result in unrealistic, unphysical results.  First, abrupt changes of
the confining potential near the $y$-edges, such as shown in the left
panel of Fig.~\ref{fig4}, result in strong, fast oscillations in the
resistances (similar to those displayed in  Fig.~\ref{fig9}(b), but with an
amplitude of order $h/e^2$).  Secondly, raising the confining potential
to near zero at the $x$-edges, as shown in the same case, also leads
to undesired  consequences, for example half quantization $R^H=R^L =
h/(2e^2)$ over a considerable energy range.  This is because leads 1
and 4 are only connected to the left-most and right-most LL states. In
this case, the energies of these states are close to zero, so they
appear to be huge barriers blocking the electrons on the lead to enter
the sample.  Electrons can only tunnel through these barriers with
very small probabilities, and no chiral currents can exit at the
$x$-edges unless $E_F \approx 0$. In other words, this is equivalent
to having very bad contacts, a situation which is avoided in the
experiments.

The right panel of Fig.~\ref{fig4} shows an improved confining
potential.  The confining potential varies smoothy near the $y$-edges,
and the $x$-edges are no longer fixed at constant value.  The
$y$-edges are smooth because in a real sample, we do expect a smooth
rise of the confining potential on the edges of the Hall bar. Its
gradient is determined by how the 2DES is defined on the sample, as
well as various screening effects.

Figure~\ref{fig9} plots the Hall resistances calculated with $V_c$
shown in Fig.~\ref{fig4}~(right).  Panel (a) shows an overview of the
data. Here $V_{\rm gap}=10$meV, so that $E_F$ was varied over a range
spanning the entire confining potential.  Particle-hole symmetry is
obvious. Panel (b) zooms in on the central region close to $E_F=0$ in
panel (a) and reveals the minute scale of oscillations in $R^H$. In
this regime, the edge states have been fully established, and the
total conductance matrix is almost exactly $2\hat{g}^{(0)}$.  The
small oscillations also exhibit the built-in particle-hole symmetry.
Panel (c) zooms in on the transition region. One can see that the
transition from first to the second plateau occurs within a small
energy interval. This is because in the absence of disorder, the edge
states are established immediately once $E_F> -V_{\rm gap}$.  In all
panels, we see that the two Hall resistances  are indeed identical.

\begin{figure}[b]
\centering \includegraphics[width=0.7\columnwidth]{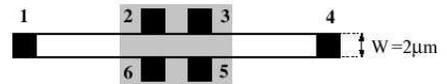}
\caption{A sketch of the sample used by Peled {\em et
        al.},\citep{Peled03a,Peled03b,Peled04} drawn to scale.  The region
        spanned by our LL Hilbert subspace is comparable to the size of
        the shaded area.}
\label{fig5}
\end{figure}

Figure~\ref{fig10} shows the longitudinal resistances for the same
simulation. All the expected symmetries are again observed. Except for
the very narrow interval of energies where the transition takes place,
the longitudinal resistances are vanishingly small. This is expected,
since for all energies except near $\pm V_{\rm gap}$ all transport is
due to edge modes, which do not contribute to $R^L$. The small
deviations from zero near $E_F=0$ are magnified in panel (b). In panel
(c), we show the peaks associated with the transition.  The symmetries
observed in the results confirm the numerical accuracy of the
simulations. We are now confident that in the absence of disorder, the
simulation shows a clear integer quantum Hall transition.  Thus, we
have confidence in attributing extra features in the full simulations
to the effects of the disorder.

As we just showed, the confining potential of Fig.~\ref{fig4}~(right)
gives physically reasonable results. However, we make one more
modification, to account for the short size of the samples we use in
numerical simulations.  Fig.~\ref{fig5} shows the geometry of the real
sample used in the experiments.\cite{Peled03a, Peled03b, Peled04} The
length of the real sample is approximately 20~$\mu$m. Given our
computational resources, we simulate the relatively short shaded area,
so that after adding the confining potential, our Hall bar has the
same thickeness and distance between the 4 central voltage terminals,
but is only 4~$\mu$m long. Thus, our current source and drain (leads 1
and 4) are much closer to the voltage probes than in experiments.

Because of this short distance, we frequently observe direct tunneling
between leads 1 or 4 and their nearest neighbor terminals (2, 6
respectively 3, 5). In the simulation, this kind of tunneling leads to
large symmetric matrix elements in the conductance matrix,
e.g. $g_{12}\approx g_{21} \sim e^2/h$, and abnormal behavior of $R^L$
and $R^H$. Experimentalists take great care to avoid direct tunneling
between leads, which constitutes a short-circuit. To avoid such direct
tunneling in the simulation, we add triangular potential barriers in
the corners of the Hall bar.  With these, the edges are effectively
prolonged, corresponding to an increased separation between the
voltage probes and current source and drain. Fig.~\ref{fig6} shows the
final form of the confining potential used in our simulations with six
terminals.  \nocite{*}

\bibliographystyle{apsrev} 
%\bibliography{article}

\begin{thebibliography}{24}
\bibitem{note} Spin-dependent scattering --although it may be very
weak -- is necessary for the relaxation of electrons to the lowest
spin-polarized LL as the magnetic field is increased.
\end{thebibliography}

\end{document}